\documentstyle[psfig]{mn}

\def\etal{{\it et al.\ }}
\def\eg{{\it e.g.\ }}

\def\spose#1{\hbox to 0pt{#1\hss}}
\def\approxlt{\mathrel{\spose{\lower 3pt\hbox{$\sim$}}
	\raise 2.0pt\hbox{$<$}}}
\def\approxgt{\mathrel{\spose{\lower 3pt\hbox{$\sim$}}
	\raise 2.0pt\hbox{$>$}}}
\def\approxpropto{\mathrel{\spose{\lower 3pt\hbox{$\sim$}}
	\raise 2.0pt\hbox{$\propto$}}}
\mathchardef\twiddle="2218

\def\multleft#1{\hbox to size{\vbox {\halign {\lft{##}\cr #1}}\hfill}\par}
\def\multright#1{\hbox to size{\vbox {\halign {\rt{##}\cr #1}}\hfill}\par}

\def\today{\ifcase\month\or January\or February\or March\or April\or May\or
      June\or July\or August\or September\or October\or November\or December\fi
      \space\number\day, \number\year}
\def\<{\thinspace}

\def\apc{\rm atom cm$^{-2}$}

\def\erg{{\rm\thinspace erg}}

\def\km{{\rm\thinspace km}}

\def\Mpc{{\rm\thinspace Mpc}}
\def\Msun{\hbox{$\rm\thinspace M_{\odot}$}}

\def\s{{\rm\thinspace s}}
\def\yr{{\rm\thinspace yr}}

\def\ergps{\hbox{$\erg\s^{-1}\,$}}

\def\kmps{\hbox{$\km\s^{-1}\,$}}

\def\Msunpyr{\hbox{$\Msun\yr^{-1}\,$}}

\def\kmpspMpc{\hbox{$\kmps\Mpc^{-1}$}}

\def\apc{\rm atom cm$^{-2}$}

\title[Chandra measurements of the distribution of mass in Abell 2390]
{Chandra measurements of the distribution of mass in the 
luminous lensing cluster Abell 2390 }  
\author[S.W. Allen et al.]
{\parbox[]{6.in} {S.W. Allen, S. Ettori and A.C. Fabian \\
\footnotesize
Institute of Astronomy, Madingley Road, Cambridge CB3 0HA
}}
\begin{document}
\maketitle
\begin{abstract}
We present spatially-resolved X-ray spectroscopy of the luminous lensing 
cluster Abell 2390, using observations made with the Chandra observatory. The 
temperature of the X-ray gas rises with increasing radius within the central 
$\sim 200$ kpc of the cluster, and then remains approximately isothermal, with
$kT =11.5^{+1.5}_{-1.6}$ keV, out to the limits of the observations at 
$r \sim 1.0$ Mpc. The total mass profile determined from the Chandra data has 
a form in good agreement with the predictions from numerical simulations. 
Using the parameterization of Navarro, Frenk \& White (1997), we measure a 
scale radius $r_{\rm s} \sim 0.9$ Mpc and a concentration parameter 
$c \sim 4$. The best-fit X-ray mass model is in good agreement with 
independent gravitational lensing results and optical measurements of the 
galaxy velocity dispersion in the cluster. The X-ray gas-to-total-mass ratio 
rises with increasing radius with $f_{\rm gas}=21\pm10$ per cent at 
$r=0.9$ Mpc. The azimuthally-averaged $0.3-7.0$ keV surface brightness 
profile exhibits a small core radius and a clear `break' at $r \sim 500$ kpc, 
where the slope changes from $S_{\rm X} \approxpropto r^{-1.5}$ to 
$S_{\rm X} \approxpropto r^{-3.6}$. The data for the central region of the 
cluster indicate the presence of a cooling flow with a mass deposition rate 
of $200-300$ \Msunpyr~and an age of $2-3$ Gyr. 
\end{abstract}


\section{Introduction}

Accurate measurements of the masses of clusters of galaxies 
are of profound importance to cosmological studies. Originally, most 
measurements of cluster masses were based on optical 
studies of their galaxy dynamics, wherein the motions of individual galaxies 
were used to trace the cluster potentials. Although such studies were 
shown to be sensitive to systematic uncertainties due to velocity 
anisotropies, substructure and projection effects (\eg Lucey 1983; 
Frenk \etal 1990; van Haarlem, Frenk \& White 1997), more recent work based on 
large galaxy samples and employing careful selection techniques, has
lead to significant progress 
(\eg Carlberg \etal 1996; den Hartog \& Katgert 1996; 
Fadda \etal 1996; Mazure \etal 1996; Borgani \etal 1999; Geller, Diaferio \& 
Kurtz 1999; Koranyi \& Geller 2000).

Recent years have also seen the development of two further 
techniques for measuring the masses of clusters, based on
X-ray observations and studies of gravitational 
lensing by clusters, respectively. X-ray mass measurements use the 
assumption that the X-ray emitting gas which pervades 
clusters is in hydrostatic equilibrium; 
the total mass distribution is determined once the radial distributions 
of the X-ray gas density and temperature are known (see \eg Sarazin 1988). 
Since the X-ray emissivity is proportional to the square of the gas 
density, and the 
relaxation timescale for the X-ray gas is relatively short (of the order 
of a few sound crossing times), the X-ray method is relatively 
free from the projection and 
substructure effects which hamper the aforementioned optical studies. 
 
In contrast to the X-ray and optical dynamical techniques, 
gravitational lensing offers a method for measuring the 
projected masses through clusters that is 
essentially free from assumptions about the dynamical state of 
the gravitating matter (see \eg Fort \& Mellier 1994; Bartelmann 
\& Schneider 1999; Mellier 1999 for reviews). 
The primary observational challenges 
of requiring deep exposures, excellent seeing conditions, 
wide field imaging and accurate point spread function models have now 
been mostly overcome with improved instrumentation (\eg Bacon \etal 2000 and 
references therein), although the recovery of the 
three-dimensional mass distributions in clusters can be complicated by 
projection effects and uncertainties in the redshift distributions of the 
lensed sources.

Clearly, the best approach when attempting to reliably measure 
the masses of clusters is to combine these three methods.
The first combined X-ray and lensing studies of galaxy clusters 
(Miralda-Escud\'e \& Babul 1995) suggested that strong lensing 
masses, measured within $r \approxlt 200$ kpc of the cluster centre, might 
typically overestimate X-ray-determined masses by a factor of 
$\sim 2-3$. This sparked debate into the possible effects of 
oblate/prolate cluster geometries, projection effects, complex temperature 
structures and pressure support from bulk and/or turbulent motions 
and magnetic fields in the X-ray gas (\eg Miralda-Escud\'e \& Babul 1995; 
Loeb \& Mao 1994; 
Waxman \& Miralda-Escud\'e 1995; Kneib \etal 1995; 
Bartelmann \& Steinmetz 1996). Later work (Allen, Fabian \& Kneib 1996; 
Allen 1998; B\"ohringer \etal 1998; Wu 2000) highlighted a clear 
difference between the results obtained for cooling-flow (hereafter CF) and 
non-cooling flow (NCF) clusters. 
For CF clusters, the X-ray and strong lensing mass measurements 
generally show good agreement, once the effects of 
the cooling flows are accounted for in the X-ray analysis. 
For NCF systems, however, the masses inferred from the strong 
lensing data invariably exceed the X-ray values, 
determined under the hydrostatic assumption, by a factor of $2-4$. 

The origin of the different results obtained for CF and NCF 
clusters is thought to lie in the different dynamical 
states of these systems: whereas X-ray and optical imaging and detailed 
lensing analyses of CF clusters show them to be relatively regular and 
dynamically-relaxed systems, NCF clusters generally 
appear to be undergoing major subcluster merger events 
(\eg Edge, Stewart \& Fabian 1992; 
Buote \& Tsai 1996; Kneib \etal 1995; Smail \etal 1995, 1997; 
Squires \etal 1997). Significant offsets between the X-ray and 
lensing centroids are observed in NCF clusters, demonstrating a 
loss of hydrostatic equilibrium in their central regions. 
The X-ray core radii for NCF systems also appear to have been inflated by 
the dynamical activity, 
in agreement with the predictions from numerical simulations 
(\eg Roetigger \etal 1996); this inflation of the X-ray core radii can 
account for the bulk of the X-ray/strong lensing mass discrepancy in most NCF 
systems (Allen 1998).  On larger spatial scales ($r \approxgt 0.5$ Mpc), 
comparisons between weak lensing, X-ray and optical dynamical 
mass measurements generally provide a consistent picture, 
with excellent agreement between the results obtained for 
both CF and NCF clusters (\eg Squires \etal 1996; Smail \etal 1997; 
Wu \& Fang 1997; Allen 1998; Lewis \etal 1999). This suggests that the 
loss of hydrostatic equilibrium in NCF clusters is primarily restricted 
to their inner regions.

The most significant uncertainty associated with the X-ray mass 
measurements in previous joint X-ray/lensing studies 
has been the absence of any direct measurements of the X-ray temperature 
profiles in the clusters, which impacts directly on the mass measurements 
through the hydrostatic equation. Although limited spatially-resolved 
spectroscopy for bright, nearby clusters was possible using ASCA and 
Beppo-SAX observations (\eg Markevitch \etal 1998; Kikuchi \etal 1999;
de Grandi \& Molendi 1999; White 2000; Irwin \& Bregman 2000), for the
more distant lensing clusters, typically observed at redshifts 
$z \approxgt 0.2$, only a single, integrated cluster spectrum was normally 
available. In most cases, only a mean emission-weighted 
X-ray temperature was therefore determined, 
although more sophisticated studies also accounted for the effects of 
cooling flows on the X-ray data (\eg Allen 1998; B\"ohringer \etal 1998). 
For the mass analyses, it was then necessarily assumed that the mass-weighted 
temperature profile followed some particular form, and usually that it 
remained approximately isothermal with radius. However, the validity of this
assumption remains uncertain, especially within the strong lensing regime.

The launch of the Chandra Observatory (Weisskopf \etal 2000) 
in 1999 July provides the first opportunity for detailed, 
spatially-resolved X-ray spectroscopy of clusters of galaxies 
at moderate redshifts. The Advanced CCD Imaging Spectrometer (ACIS) 
on Chandra permits the first direct, simultaneous measurements of the X-ray 
temperature and density profiles and, via the hydrostatic assumption, the
mass distribution in luminous lensing clusters, spanning both the weak and 
strong lensing regimes. In this paper we present the first results from 
Chandra observations of the massive CF lensing cluster Abell 2390, which 
has been the subject of several previous combined X-ray/optical/lensing 
studies (\eg Piere \etal 1996, Squires \etal 1996, 
Allen 1998, B\"ohringer \etal 1998, Lewis \etal 1999). We present 
detailed results on the mass distribution in the cluster determined from the 
Chandra data and compare our results with those from detailed strong and 
weak lensing analyses and optical dynamical studies. We also re-examine the 
properties of the cooling flow in the cluster and discuss 
their relation to the dynamical history of the system.

The cosmological parameters $H_0$=50 \kmpspMpc, $\Omega = 1$ and 
$\Lambda = 0$ are assumed throughout. At the redshift of Abell 2390 
($z=0.2301$) an angular scale of 1 arcsec corresponds to a physical length 
of 4.652 kpc.

\section{Observations}

The Chandra observations of Abell 2390 were carried out using the 
ACIS on 1999 November 7. The target was observed in the 
back-illuminated CCD detectors, close to the nominal aim point for the 
ACIS S3 detector. (The source was positioned near the centre of node-1 
on the chip). The focal plane temperature at the time of the observations 
was -110C.

We have used the  CIAO software (version 1.1.3) and the level-2 events file 
provided by the standard Chandra pipeline
processing for our analysis. The light curve for the observation was 
of high quality with no strong background flaring.
Only those X-ray events with grade classifications of 0,2,3,4 and 6 
were included in our
final cleaned data set, for which the net exposure time was 9.13ks.

\section{X-ray imaging analysis}

\subsection{X-ray morphology}

\begin{figure*}
\vspace{2cm}
\hbox{
\hspace{0.2cm}\psfig{figure=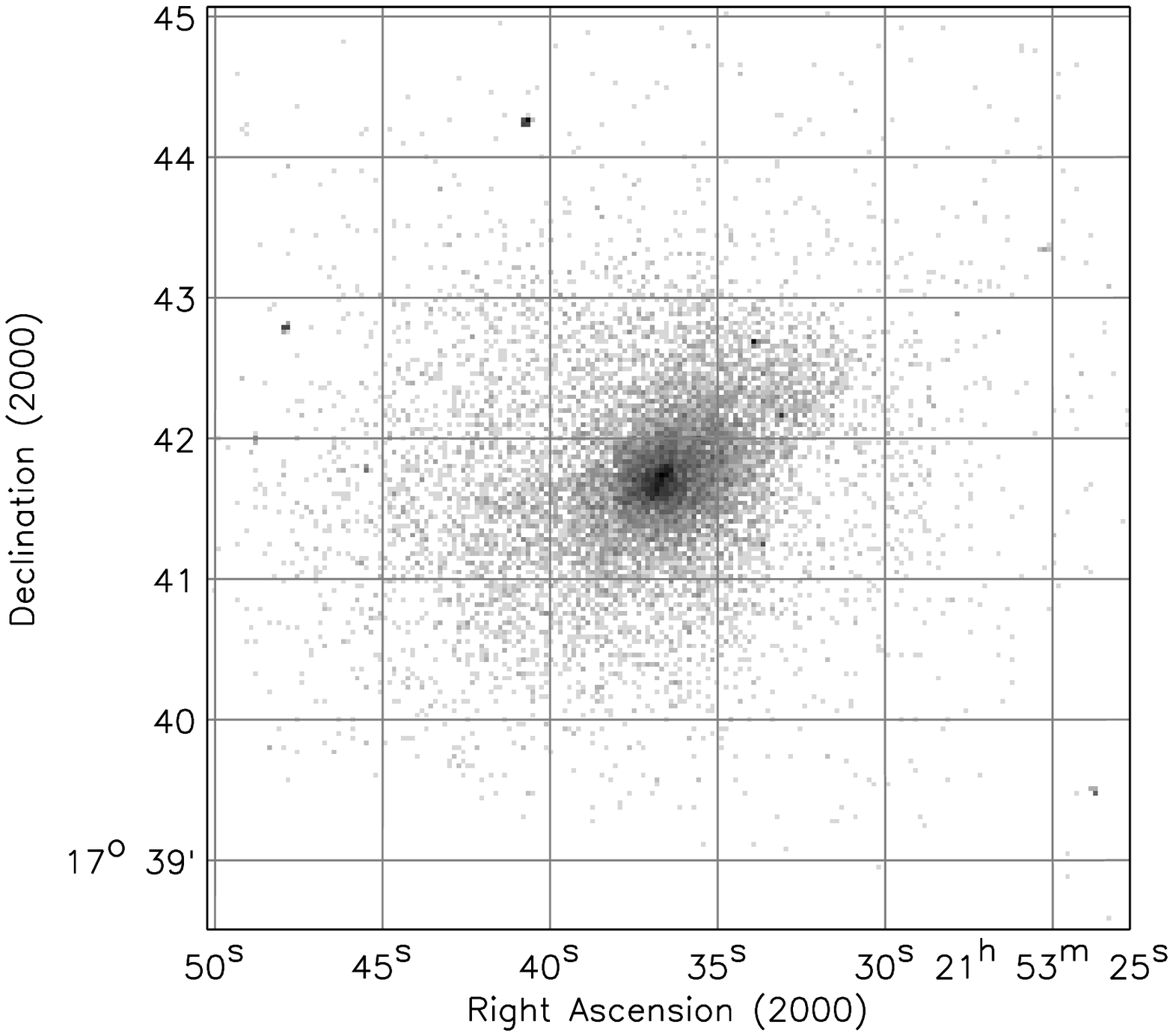,width=0.45 \textwidth,angle=0}
\hspace{0.8cm}\psfig{figure=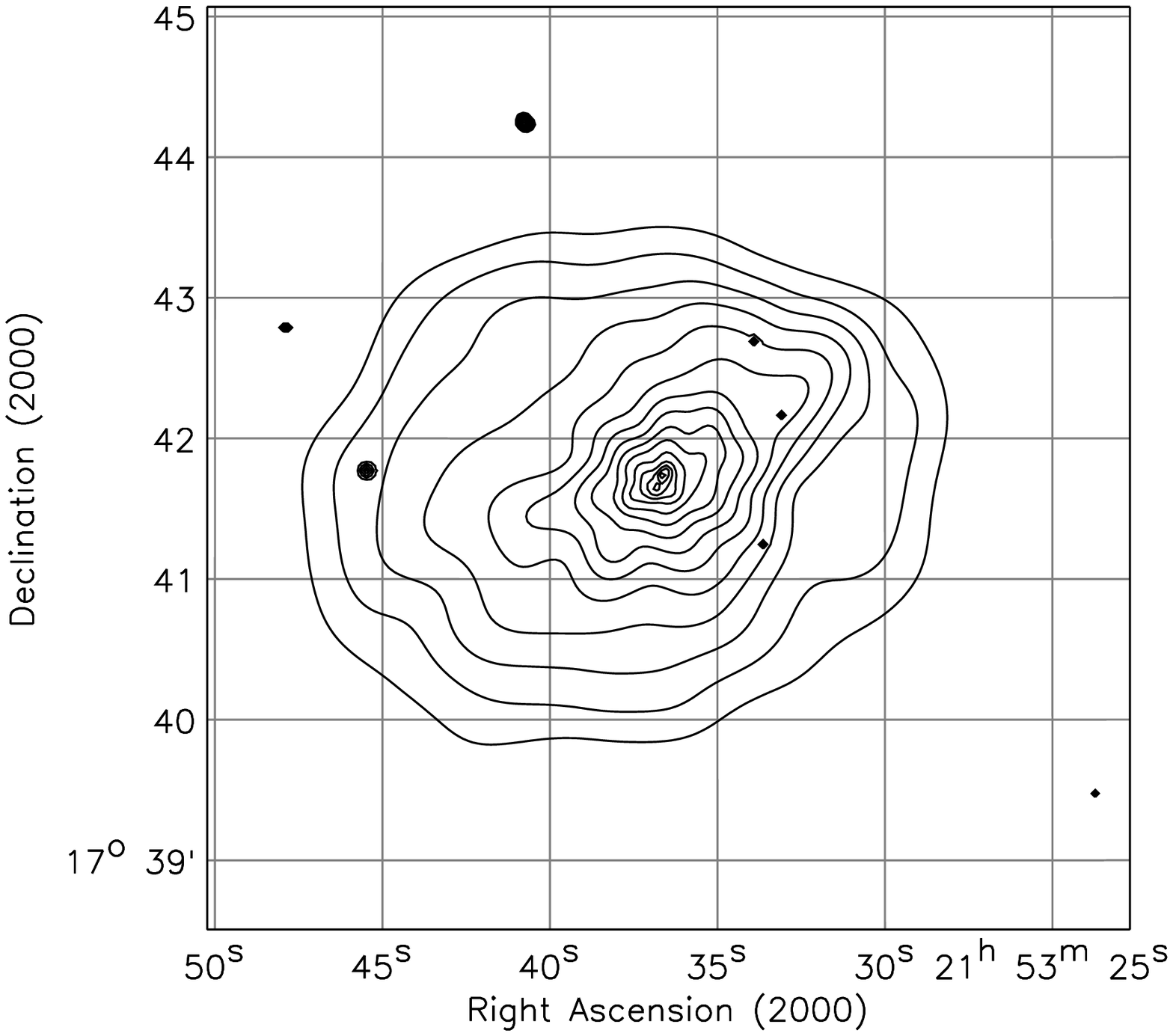,width=0.45 \textwidth,angle=0}
}
\caption{(Left panel) The raw $0.3-7.0$ keV Chandra image of Abell 2390. 
The pixel size is 4 detector pixels (1.97 arcsec). (Right panel) Contour 
plot of the same region, adaptively smoothed using the code of Ebeling, 
White \& Rangarajan (2000), with a 
threshold value of $3.5 \sigma$. The contours have equal logarithmic spacing. }\label{fig:im1} 
\end{figure*}

The raw $0.3-7.0$ keV image of the central $6 \times 6$ 
arcmin$^2$ ($1.8 \times 1.8$ Mpc$^2$) region of Abell 2390 is 
shown in Fig. \ref{fig:im1}(a). The pixel size is 
$1.97  \times 1.97$ arcsec$^2$, corresponding to  $4 \times 4$ raw detector 
pixels. Fig. \ref{fig:im1}(b) shows an adaptively smoothed version of the same 
image, using the smoothing algorithm of Ebeling, White \& Rangarajan  
(2000). The position of the peak of the X-ray emission from the cluster, 
21h53m36.77 +17d41m42.8s 
(J2000.), is in excellent agreement with the optical centroid for the dominant 
cluster galaxy of 21h53m36.76 +17d41m42.9s (J2000.; Pierre \etal 1996). The 
X-ray image is elongated along an approximately northwest-southeast 
direction, in a similar manner to the optical isophotes of the dominant 
galaxy (Pierre \etal 1996).

\begin{figure}
\vspace{2cm}
\hbox{
\hspace{0.0cm}\psfig{figure=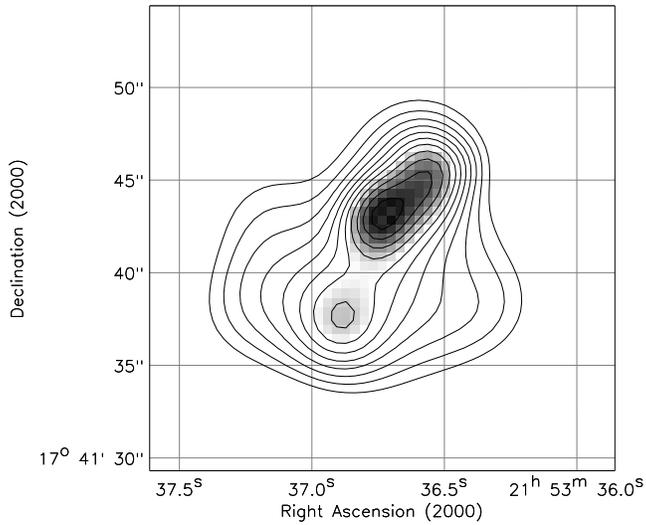,width=0.45 \textwidth,angle=0}
}
\caption{An adaptively smoothed $0.3-7.0$ keV image of the central regions of 
Abell 2390 on a finer spatial scale (pixel size $0.492 \times 0.492$ 
arcsec$^2$, which is equivalent to $1\times1$ raw detector pixels). Note the 
ridge of enhanced emission extending $\sim 3-4$ arcsec to the northwest of 
the X-ray peak which is coincident with the excess blue continuum and optical/UV line 
emission reported by L\'emonon \etal (1998) and Hutchings \& Balogh (2000).
}\label{fig:im2} 
\end{figure}

Some substructure is apparent in the Chandra image. In the central region, 
a relatively bright ridge of enhanced emission extends $\sim 
3-4$ arcsec to the northwest of the X-ray peak (Fig. \ref{fig:im2}). 
Enhanced emission in the same direction is also observed in the optical 
blue continuum and optical/UV line emission from the dominant cluster 
galaxy (L\'emonon \etal 1998; Edge \etal 1999; 
Hutchings \& Balogh 2000). A correspondence between the brightest X-ray 
emission with excess blue continuum and optical/UV line emission 
(probably associated with the formation of young, massive stars) is 
also observed in the nearby CF clusters Hydra-A (McNamara \etal 2000) and 
Abell 1795 (Fabian \etal 2000b). A second, fainter region of 
enhanced X-ray emission extends $\sim 6$ arcsec to the south-southeast. 

On medium ($r \sim 40-80$ arcsec) scales, the X-ray emission is extended 
to the west-northwest, in a similar manner to the optical luminosity 
distribution 
and lensing mass models of Piere \etal (1996; these authors also note the 
presence of an X-ray extension in the same direction using ROSAT High 
Resolution Imager data). 
On large scales ($\approxgt 2$ arcmin) the X-ray emission is extended 
towards the east. The presence of such substructure in the Chandra image 
suggests that the cluster has not fully relaxed following it's most recent 
merger activity. However, the agreement between the X-ray, optical and 
gravitational lensing mass measurements discussed in Section 5.2 argues 
that overall the assumption of hydrostatic equilibrium in the 
cluster is a reasonable one. 

Three point sources are detected to the west of the X-ray peak, in the central 
regions of the cluster, at positions 21h53m33.19 +17d42m08.4s, 
21h53m33.74 +17d41m13.3s, and 21h53m34.02 +17d42m39.9s. 
The X-ray and sub-mm properties of these sources are discussed by 
Fabian \etal (2000a). Four more point sources are also visible 
at larger radii in Fig. \ref{fig:im1}(b).

\subsection{The surface brightness profile}

\begin{figure*}
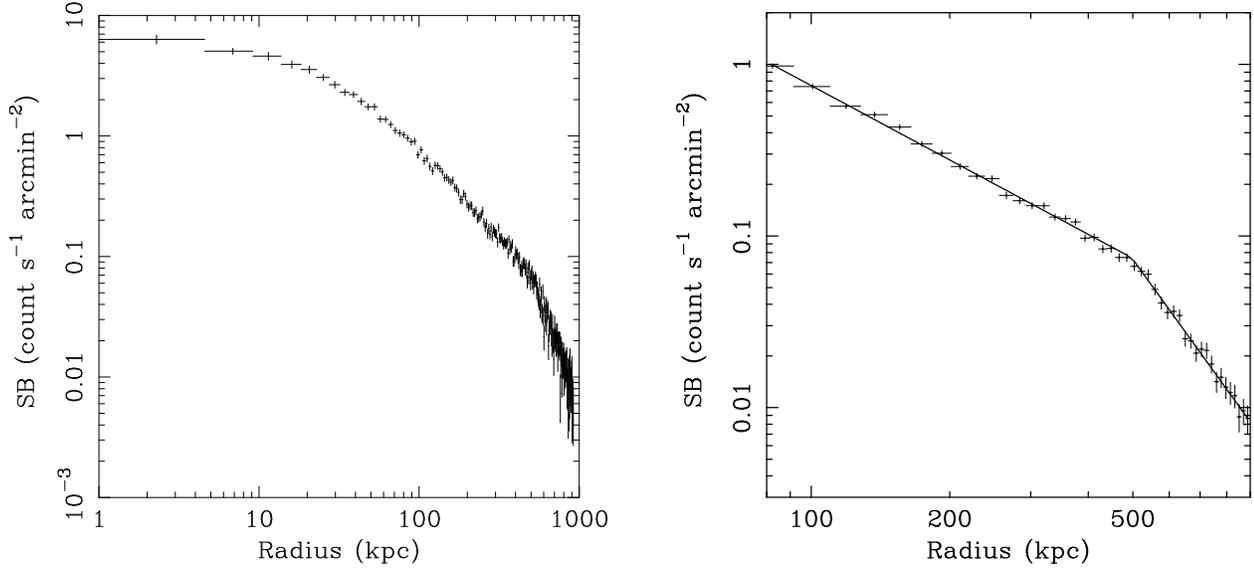

\vspace{2cm}
\hbox{
\hspace{0.2cm}\psfig{figure=sb_0pt3to7_fuf_kpc_pretty_surbri3.ps,width=0.45 \textwidth,angle=270}
\hspace{0.8cm}\psfig{figure=sb_0pt3to7_fuf_4asec_kpc_pretty_new2.ps,width=0.433 \textwidth,angle=270}}
\caption{(a) The background-subtracted, flat-fielded, azimuthally-averaged 
radial surface brightness profile for Abell 2390 in the $0.3-7.0$ keV band.
The binsize is 0.984 arcsec (4.65 kpc). (b) The same profile, rebinned by a 
factor of 4, in the range $80-900$ kpc, with the best-fitting 
broken power-law model overlaid. At the break radius of 
$491^{+16}_{-7}$ kpc, the slope of the surface brightness profile 
($S_{\rm X} 
\propto r^{\alpha}$) changes from $\alpha = -1.46 \pm 0.02$ to 
$\alpha = -3.60 \pm 0.17$.}\label{fig:surbri}
\end{figure*}

The azimuthally-averaged, $0.3-7.0$ keV X-ray surface 
brightness profile for Abell 2390 is shown in Fig. \ref{fig:surbri}. 
The profile has been flat-fielded and background subtracted using 
a rectangular background region of size $400 \times 50$ arcsec$^2$, 
located $\sim 5$ arcmin from the cluster centre. 
All obvious point sources were excluded from the analysis. 

The X-ray emission from the cluster extends beyond the 5 
arcmin ($\sim 1.4$ Mpc) radius associated with our on-chip 
background region (see also Pierre \etal 1996; 
B\"ohringer \etal 1998). However, beyond this radius 
background counts dominate the flux in the ACIS-S3 detector.
Fig. \ref{fig:surbri} shows the data for the central 900 kpc ($193$ arcsec), 
for which systematic errors associated with the 
background subtraction and flat fielding are negligible.
The bin-size in Fig. \ref{fig:surbri}(a) is 2 detector pixels 
(0.984 arcsec). Fig. \ref{fig:surbri}(b) shows the data for the outer 
regions of the cluster with a larger binsize of 8 
detector pixels (3.94 arcsec).

Within a radius of 500 kpc ($107$ arcsec), the X-ray surface 
brightness profile can be parameterized ($\chi^2 = 119$ for 
106 degrees of freedom) by a standard $\beta$-model 
(\eg Jones \& Forman 1984) of the form $S(r) = 
S(0)\left [ {(1+{r/r_{\rm c})}^2} \right ]^{1/2-3\beta}$, 
with a core radius $r_{\rm c} = 28.7\pm1.3$ kpc and a 
slope parameter $\beta =0.412\pm0.003$ ($1 \sigma$ errors; $\Delta 
\chi^2=1.0$). On larger scales, however, the $\beta$-model 
does not provide an acceptable fit: examining the data for the central 
900 kpc radius, we obtain $\chi^2 = 909$ for 194 degrees of 
freedom, with best-fit parameter values of $r_{\rm c} = 55$ kpc and 
$\beta =0.48$). Ignoring the central $r \sim 80$ kpc (18 arcsec) region, 
associated with the possible cooling flow (Section 6), the fit is improved 
($\chi^2 = 485$ for 177 degrees of freedom, with best-fit parameter 
values $r_{\rm c} = 160$kpc and $\beta =0.58$) although is still formally 
unacceptable. 

\begin{figure}
\vspace{2cm}
\hbox{
\hspace{0.0cm}\psfig{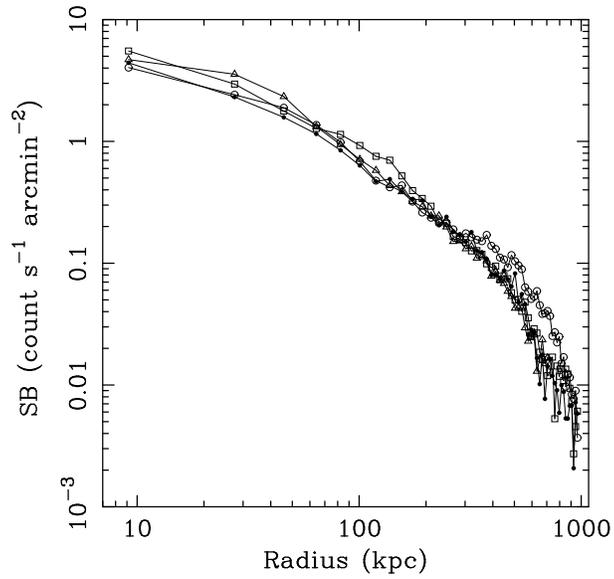}
}
\caption{The $0.3-7.0$ keV surface brightness profile in quadrants 
covering position angles $315-45$ degrees (filled circles), $45-135$ degrees 
(open circles), $135-225$ degrees (triangles) and $225-315$ degrees (squares).
The extension towards the East is also evident in Fig. \ref{fig:im1}. The 
results for the other  
quadrants are in good agreement with each other.}\label{fig:surbri_azi}
\end{figure}

The main reason for the poor fit obtained with the $\beta$-model 
at larger radii is the presence of a `break' in the surface 
brightness profile at $r \sim 500$ kpc (Fig. \ref{fig:surbri}b). 
This break is not obviously due to substructure in any particular 
direction in the cluster:
Fig. \ref{fig:surbri_azi} shows the surface brightness profile measured
in the four quadrants covering position angles 
$45-135$, $135-225$, $225-315$ and $315-45$ degrees. We see that the profile 
appears remarkably similar in three of the four directions, although the 
emission is slightly more extended towards the east (as is also evident 
in the images presented in Fig. \ref{fig:im1}).

A good fit to the surface brightness profile beyond the central 
cooling region can be obtained using a simple broken power-law model. 
Fitting the data from $80-900$ kpc, we obtain $\chi^2 = 173$ for 176 
degrees of 
freedom, with a break at a radius of $491^{+16}_{-7}$ kpc, and slopes in 
the regions internal and external to the break radius of $-1.46\pm0.02$ and 
$-3.60 \pm 0.17$, respectively. Interestingly, these slopes 
are similar to the values expected at small and 
large radii for the dark matter in a Navarro, Frenk \& White 
(1997; hereafter NFW) potential in which $\rho \propto 1/((r/r_{\rm s})(1+r/{\rm r_s})^2$). However, isothermal gas in an NFW-like potential should not 
exhibit a sharp break at the scale radius $r_{\rm s}$, but rather a slow 
rollover  (although individual clusters in the simulations 
presented by Thomas \etal 2000 do exhibit sharp breaks in their dark matter 
distributions). Fitting the surface brightness profile external to the 
cooling flow ($80-900$ kpc) with the prescription for isothermal gas in 
an NFW potential 
described by Ettori \& Fabian (1999), we measure $r_{\rm s} = 
610^{+60}_{-50}$ kpc (formal $1\sigma$ errors), with $\chi^2 = 420$ for 
177 degrees of freedom. 
Thus, the fit with the NFW mass model assuming strict isothermality in 
the X-ray gas is formally unacceptable, although provides a better 
description of the data in the $80-900$ kpc region than the  
$\beta$-model. As discussed in Section 4, the Chandra data show that 
the X-ray gas in Abell 2390 is not isothermal and that the temperature 
rises with increasing radius within the central $r \sim 200$ kpc. 
In Section 5 we show that an NFW mass model can provide a good 
description of the Chandra data, once the assumption of isothermality is 
relaxed.

\section{Spatially-resolved spectroscopy}

\subsection{Method of analysis}

For our spectral analysis, we divided the cluster 
into annular regions, as detailed in 
Table 1. A spectrum was extracted from each region in 
1024 Pulse Height Analyser (PHA) channels. The spectra were  
re-grouped to contain a minimum of 20 counts
per PHA channel, thereby allowing $\chi^2$ statistics to be used. (For the
two outer annuli, 
a larger grouping of 40 counts per PHA channel was used, due to
the increased background contribution). Background spectra, appropriate 
for the regions studied, were extracted from the 
ACIS-S3 blank-field data sets available from 
the Chandra X-ray Center. All obvious point sources were 
masked out and excluded from the analysis. 
Separate photon-weighted response matrices and 
effective area files were constructed for each region 
using the calibration and response files appropriate 
for the focal plane temperature, available from the Chandra
X-ray Centre. 

Two separate energy ranges were examined. 
Firstly, a conservative $0.5-7.0$ keV band was used
over which the calibration of the back-illuminated 
CCD detectors is currently best understood. Secondly, 
for the central 100 kpc region, where a cooling flow is 
thought to exist (Section 6), we 
have also examined a more extended $0.3-7.0$ keV energy range, 
which provides extra constraints on the 
presence of cool emission components and/or  
intrinsic absorption in the cluster.

\subsection{The spectral models}

The analysis of the spectral data has been 
carried out using the XSPEC software package 
(version 11.01; Arnaud 1996). The spectra 
were modeled using the 
plasma emission code of Kaastra \& Mewe
(1993; incorporating the Fe L calculations of Liedhal, Osterheld \&
Goldstein 1995) and the photoelectric absorption models of 
Balucinska-Church \& McCammon (1992). We first examined each annular 
spectrum using a simple, single-temperature model with the absorbing column
density fixed at the nominal Galactic value ($N_{\rm H} = 6.8 \times
10^{20}$\apc; Dickey \& Lockman 1990). This model is hereafter
referred to as model A. The free parameters in model A were the 
temperature ($kT$) and metallicity ($Z$) of the plasma 
(measured relative to the solar photospheric values of Anders 
\& Grevesse 1989, with the various elements assumed to be present 
in their solar ratios) and the emission measure ($K$). We also examined 
a second single-temperature model (model B) which was 
identical to model A but with the absorbing column 
density $(N_{\rm H})$ also included as a free parameter in the fits.  

The image deprojection and X-ray colour profile analyses discussed in 
Section 6 indicate the presence of a strong cooling flow in the central
$r \approxlt 100$ kpc region of the cluster. 
We have therefore also examined the 
spectral data for this region using a series of more sophisticated, 
multiphase models in which the emission properties of the cooling flow 
were explicitly accounted for. The first such 
model (model C1) introduced an extra emission component into model A, 
with a spectrum appropriate for gas cooling at constant pressure from
 the ambient cluster temperature (following the prescription of Johnstone 
\etal 1992). The normalization of this component was parameterized in terms
of a mass deposition rate, ${\dot M}$, which was a free parameter in the 
fits. In the second case, model C2, the cooling gas was modelled 
as an isothermal cooling flow, following Nulsen (1998; we assume a 
value for $\eta =1$, where the integrated mass 
deposition rate within radius r, ${\dot M} \propto r^{\eta}$). 
The mean gas temperature and metallicity in both the cooling flow 
and isothermal emission component (which accounts for the emission 
from gas at larger radii viewed in projection) were assumed to be equal. 
Finally, we also examined a more general emission model,
model D, in which the cooling gas was modelled by a second, cooler
isothermal emission component, with the temperature and 
normalization of this component included as free fit parameters. 
Model D provides a more flexible parameterization, with an 
additional degree of freedom over models C1 and C2, and 
invariably provides a good match to the more specific 
cooling-flow models at the spectral resolution and 
signal-to-noise ratios typical of ACIS cluster observations.
However, the parameter values determined with model D were 
not well constrained for Abell 2390, and thus we do not quote explicit 
results for this model here.
 
With each of the cooling-flow emission models, we have also examined the 
effects of including extra absorption, 
using a variety of different absorption models. In the first
case (absorption model i), the only absorption included
was that due to cold gas in our Galaxy, with the equivalent 
column density fixed to the nominal Galactic value 
(Dickey \& Lockman 1990; For a single temperature emission model, 
this is identical to spectral model A). In the second
model (model ii), the absorption was again assumed to be 
due to Galactic (zero redshift) cold gas, 
but with the column density, $N_{\rm H}$, included as a free 
parameter in the fits. (For a single temperature emission model, 
this would be equivalent to spectral model B.) In 
the third case (absorption model iii), 
an intrinsic absorption component with column density,
 $\Delta N_{\rm H}$, due 
to cold gas at the redshift of the cluster was introduced. 
The absorber was assumed to lie in a uniform screen in front of the
cooling flow, with the column density included a free fit parameter.  
In the fourth case (model iv), the intrinsic absorption was assumed to
cover only a fraction, $f$,  of the emission from the cooling flow. 
The fifth and final absorption model (model v) was similar to model (iii) 
but with the gaseous absorber replaced by an intrinsic absorption 
edge, with the edge depth, $\tau$, and energy, $E_{\rm edge}$, 
free parameters in the fits. This more general absorption model may be used to 
approximate the effects of a dusty and/or ionized absorber.

In those cases where the absorption has been quantified in terms of  
an equivalent hydrogen column density, solar metallicity in the 
absorbing gas is assumed.
We note that in absorption
models (iii--v), the absorption acting
on the ambient cluster emission was fixed at the nominal Galactic
value. However, allowing the Galactic absorption to vary from this value did
not significantly improve the fits.

\subsection{Results from the single-phase analysis}

The best-fit parameter values and $1\sigma$ ($\Delta \chi^2 = 1.0$) 
and 90 per cent ($\Delta \chi^2 = 2.71$) confidence limits determined from 
the fits in the $0.5-7.0$ keV band with the single temperature models 
are summarized in Tables 1 and 2. 
The temperature profile determined with spectral model A is shown in 
Fig. \ref{fig:kt}(a). The measured temperature 
is approximately isothermal beyond a 
radius of 200 kpc, out to the limits of the data at $r \sim 1$ Mpc. 
A combined fit to the data in the $0.2-1.0$ Mpc range with 
model A gives a mean temperature of 
$11.5^{+1.6}_{-1.3}$ keV. Fitting the $0.2-1.0$ Mpc results with a 
simple power-law model of 
the form $kT \propto r^{\alpha}$ we measure $\alpha=0.0\pm0.2$ 
($1\sigma$ bootstrap errors obtained using the 
Akritas \& Bershady 1996 modification 
of the ordinary least squares statistic.) 
We observe a clear drop in the emission 
weighted temperature within the central 100 kpc, with a 
value for the central $r = 50$ kpc of $5.58^{+0.52}_{-0.42}$ keV.

\begin{figure*}
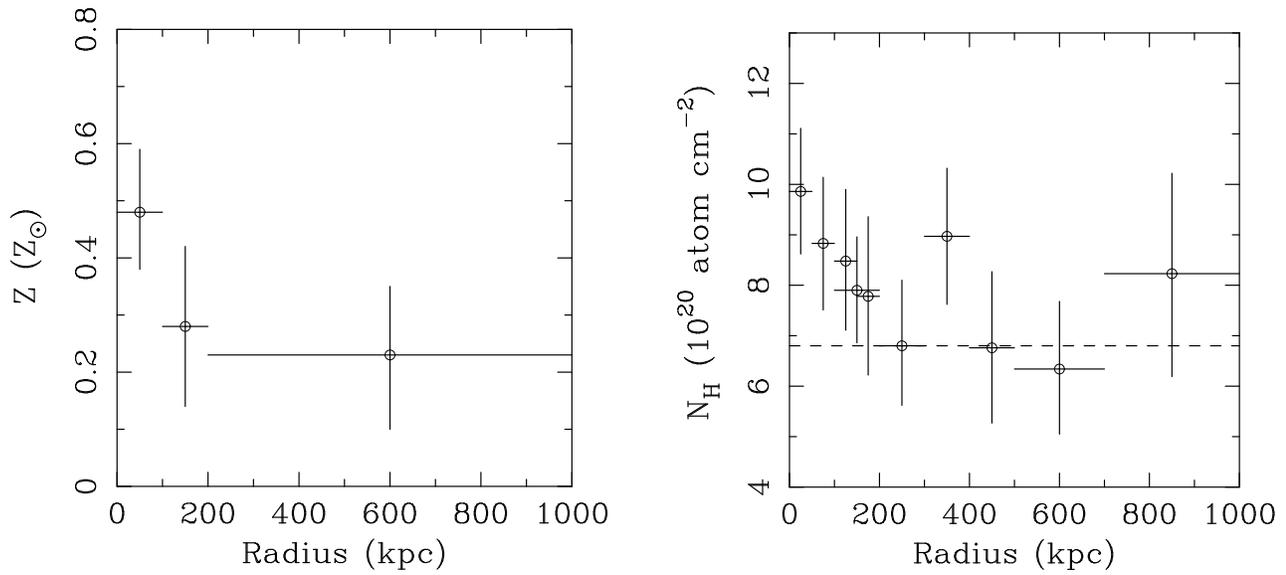

\vspace{2cm}
\hbox{
\hspace{0.2cm}\psfig{figure=abun_modela_1sigma.ps,width=0.45 \textwidth,angle=270}
\hspace{0.8cm}\psfig{figure=nh_modelb_1sigma.ps,width=0.45 \textwidth,angle=270}
}
\caption{(a) The (projected) variation of metallicity in the cluster measured with spectral 
model A in the $0.5-7.0$ keV energy range. (b) The absorbing column 
density measured with spectral model B. The dashed line shows the Galactic 
column density determined from HI studies (Dickey \& Lockman 1990).}\label{fig:abnh}
\end{figure*}

We detect marginal evidence for a metallicity gradient in the cluster, 
with a mean, best-fit value for the central 100 kpc of 
$Z=0.48^{+0.11}_{-0.10}Z_{\sun}$, which compares to a mean value 
of $0.23^{+0.12}_{-0.13}Z_{\sun}$ for the outer $0.2-1.0$ Mpc region 
(Fig. \ref{fig:abnh}a; $1\sigma$ errors). 
The results determined with spectral model B 
in the $0.5-7.0$ keV band also provide marginal evidence for increased 
absorption towards the cluster core (Fig. \ref{fig:abnh}b). 

In all cases, the $\chi^2$ values 
obtained from the fits to the $0.5-7.0$ keV data with 
the single-temperature models 
are acceptable, the only marginal case being the data for
the central 50 kpc region, where the spectrum is 
complicated by the effects of the cooling flow. 

\subsection{Multiphase analysis of the cooling core}

\begin{figure*}
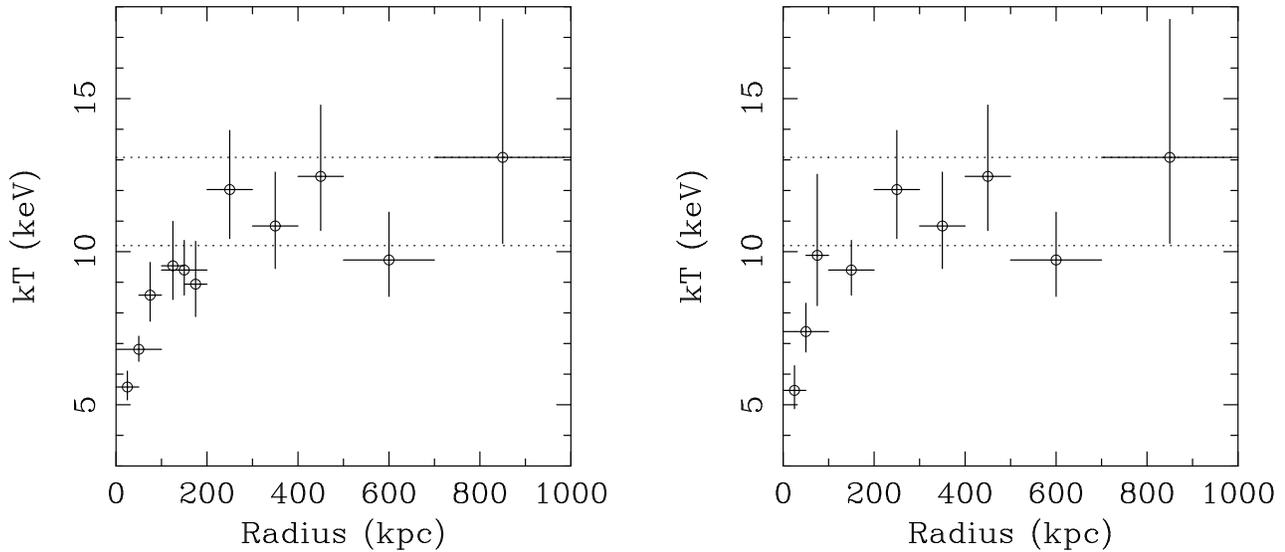

\vspace{2cm}
\hbox{
\hspace{0.2cm}\psfig{figure=kt_modela_ver2_1sigma.ps,width=0.45 \textwidth,angle=270}
\hspace{0.8cm}\psfig{figure=kt_modelc2_1sigma_final.ps,width=0.45 \textwidth,angle=270}
}
\caption{(a) The projected X-ray gas temperature profile 
(and $1\sigma$ errors) 
measured with spectral model A in the $0.5-7.0$ keV energy range. 
The dotted lines mark the 90 per cent confidence limits on the mean 
ambient cluster temperature determined from a joint analysis of the data 
in the $0.2-1.0$ Mpc region. (b) The ambient X-ray gas temperature as a 
function of radius, corrected for the effects of the cooling flow in the 
central 100kpc using spectral model C2(iii; see Section 4.4). 
For $r>100$ kpc, the values are 
determined in an identical manner to those 
in (a).}\label{fig:kt}
\end{figure*}

The results determined from the more detailed, multiphase 
analysis of the central 100 kpc radius in the extended 
$0.3-7.0$ keV energy band are summarized in Table 3. The results 
demonstrate a clear requirement for excess absorption in this 
region, over and above the nominal Galactic value 
(Dickey \& Lockman 1990), using each of the different emission models. 
The systematic variations between the column density measurements obtained 
with the different emission and absorption models are similar to those 
determined from previous ASCA studies (\eg Allen 2000). Unfortunately, 
the present Chandra data for Abell 2390 cannot statistically discriminate 
between the single-phase and multiphase emission models for the central 
100 kpc region, which provide comparable $\chi^2$ values with each 
absorption model. (We note, however, that at some level the spectrum for 
the central 100 kpc must be multiphase simply due projection effects, 
given the results on the temperature profile shown in Fig. \ref{fig:kt}(a).)

For our preferred cooling-flow emission models including 
intrinsic absorption, which provide the most consistent 
physical description of the spectral and imaging data 
for the central 100 kpc region (Section 6), we measure an integrated 
mass deposition rate ${\dot M} = 200-300$ \Msunpyr.
Fig. \ref{fig:kt}(b) shows the projected temperature profile for 
Abell 2390, corrected for the effects of the cooling flow using spectral 
model C2(iii). (For $r > 100$ kpc the results determined with spectral 
model A in the $0.5-7.0$ keV band have been used). We see that 
correcting for the effects of the cooling flow does not have a major 
impact on the temperature profile measured in the central regions 
of the cluster. Assuming that the isothermal cooling flow model 
provides a reasonable description of the data, the drop in the 
central temperature shown in Figs. \ref{fig:kt}(a,b) should then 
reflect the mass distribution in the cluster core.

The data for Abell 2390 do not provide firm constraints on the 
nature of the intrinsic absorption in the cluster. Using 
the partial-covering absorption model (iv), we find that 
high covering fractions are preferred. For the edge-like 
absorption model (v), the lower limit on the 
edge energy is essentially unconstrained: for emission models 
C1(v) and C2(v), however, the upper limit on the edge energy is 
inconsistent with the OI\,K edge of oxygen ($E_{\rm edge} \sim 0.54$ keV), 
suggesting that the absorption is unlikely to be due to oxygen 
rich dust grains (\eg Arnaud \& Mushotzky 1998; Allen \etal 2000b) or a 
warm, ionized absorber (Buote 2000). 

Assuming that the intrinsic absorption is due to cold gas lying in a uniform 
screen in front of the cooling flow, using spectral model 
C2(iii), we measure $\Delta N_{\rm H} = 2.2^{+1.1}_{-0.8} 
\times 10^{21}$ \apc. The mass of absorbing gas implied by this model 
is then $M_{\rm abs} \sim 3 \times 10^7 r_{\rm abs}^2 \Delta N_{\rm H} \Msun$, 
where $r_{\rm abs}$ is the radial extent of the absorber in 
kpc and $\Delta N_{\rm H}$ is the equivalent hydrogen column density 
in units of $10^{21}$ \apc. For $r_{\rm abs} \sim 50$ kpc (Section 6.2) 
we obtain $M_{\rm abs} \sim 1.7^{+0.8}_{-0.6}\times 10^{11}$ \Msun~( 
although Allen \& Fabian 1997 and Wise \& Sarazin 2000 argue that for 
a geometry in which the absorbing material is distributed throughout 
the X-ray emitting region, the true mass is likely to be a few times 
higher). This mass is in reasonable 
agreement with the mass expected to have been accumulated 
by the cooling flow within the same radius over its lifetime; $\sim 
{\dot M}t/2 \sim 2.8\pm1.4 \times 10^{11}$ \Msun~(the factor two in the 
denominator arises from the assumption that the integrated mass deposition 
rate, ${\dot M}$, grows approximately linearly with time).

\subsection{Spectral deprojection analysis}

The results discussed in the previous subsections are based on
the analysis of projected spectra. In order to determine 
the effects of projection on the spectral results, we have also 
carried out a simple deprojection analysis of the Chandra spectral data.

For this analysis we have used the same annular regions and have 
assumed that the emission from each spherical shell (the shells are 
defined by the same inner and outer radii as the annular regions) is 
isothermal and absorbed by the Galactic column density (spectral model A). 
The fit to the outermost 
annulus is used to determine the temperature and emission measure in the 
outermost spherical shell. The contribution from that shell to each inner 
annulus is then determined by purely geometric factors (\eg Kriss, Cioffi \& 
Canizares 1983). The fit to the second annulus inward is used 
to determine the parameters for the second spherical shell, and so forth, 
working inwards. Thus, the spectral model for the $i$th annulus working 
inwards is the weighted sum of $i$ absorbed, isothermal models. This 
parallels the usual image deprojection procedure (\eg Fabian \etal 1981).

We have used the XSPEC code and Chandra data in the $0.5-7.0$ keV band. 
The data for all eight annular spectra were fitted simultaneously in 
order to correctly determine the parameter values and confidence limits. 
The spectral model used therefore has $2n+1$ free parameters 
(where $n$ is the total number of annuli), corresponding to the temperature 
and emission measure in each spherical shell and the overall emission-weighted
metallicity (the metallicity is linked to the same value at all 
radii). Note that we have have not attempted to correct for residual emission 
from gas at radii beyond the outermost annulus since the steeply rising 
surface brightness profile of the cluster causes this emission to have 
a negligible affect on the results.  

The temperature profile determined with the spectral deprojection 
method is shown in Fig. \ref{fig:kt_xspec3d}.

\subsection{Comparison with previous work}

The mean ambient temperature for Abell 2390 of $kT = 11.5^{+1.6}_{-1.3}$ 
keV, determined from the combined analysis of the data in the $0.2-1$ Mpc 
range, is in good agreement with the previous results of $kT = 
14.5^{+15.5}_{-5.2}$keV from Allen (1998) based on ASCA observations, and 
$kT = 11.1^{+1.5}_{-1.6}$ keV from B\"ohringer \etal (1998), based on a 
joint analysis of ASCA and ROSAT Position Sensitive Proportional Counter 
(PSPC) data. (Both Allen 1998 and B\"ohringer \etal 1998 accounted for 
the effects of the central cooling flow in their modelling of the integrated 
cluster spectra. B\"ohringer \etal (1998) also examined a simpler, 
single-temperature emission model with which they measured a lower mean 
emission-weighted temperature of $kT \sim 9$ keV.) The ambient temperature 
for Abell 2390 
measured with Chandra is also in good
agreement with the predicted value of $\sim 12.0$ keV using the 
cooling-flow corrected $kT_{\rm X}/L_{\rm Bol}$ relation of Allen 
\& Fabian (1998; we assume a bolometric luminosity of $\sim 1.0 \times 
10^{46}$ \ergps~as measured by ASCA since the Chandra observations 
do not cover the whole of the cluster).

The best-fit mass deposition rate from the cooling flow of ${\dot M} 
\sim 200-300$ \Msunpyr, determined from the Chandra spectrum for the 
central 100 kpc region, is lower than previous measurements based on the 
analysis of integrated spectra for the whole cluster from ASCA 
(${\dot M} = 1500^{+600}_{-1100}$ \Msunpyr; Allen 2000), 
joint ASCA/ROSAT (${\dot M} = 700^{+150}_{-300}$  \Msunpyr; B\"ohringer 
\etal 1998) and Beppo-SAX (${\dot M} = 700^{+400}_{-300}$ \Msunpyr; 
Ettori, Allen \& Fabian 2000) observations (although the results are 
marginally consistent at the 
$\sim 95$ per cent confidence level). In part, this difference is 
likely to be due to the fact that the ambient gas temperature in the 
centre of the cluster, corrected for the effects of the cooling flow, 
is lower than the mean value measured at larger radii 
(Fig. \ref{fig:kt}b). If this drop in the central ambient temperature 
is not accounted for (as was the case in the previous ASCA and 
BeppoSAX studies, which could not spatially resolve the cooling flow 
from the hotter, outer cluster gas) then the cooler, ambient gas 
in the cluster core will also tend to be modelled as part of the cooling 
flow, and the total mass deposition rate will be overestimated. This effect is 
illustrated by the fact that a fit with spectral model C2(iii) to 
a single Chandra spectrum covering the entire central 500 kpc (radius) of 
the cluster, gives ${\dot M} = 621^{+121}_{-113}$ \Msunpyr, 
$\Delta N_{\rm H} = 1.6^{+0.8}_{-0.6} \times 10^{21}$\apc~and 
$kT = 11.9^{+4.2}_{-1.9}$ keV, in good agreement with the previous ASCA, 
ROSAT and BeppoSAX results. \footnote{Note that although the fit with 
spectral model C2(iii) to the Chandra spectrum for the 
central 500 kpc (radius) region overestimates the mass deposition 
rate from the cooling flow, the measured temperature is in good agreement 
with the true value at large radii in the cluster (Fig. \ref{fig:kt}b). 
Thus, previous ASCA and Beppo-SAX studies which attempted to account for 
the effects of cooling flows on integrated cluster spectra 
(\eg Allen 1998; B\"ohringer \etal 1998) may have overestimated the mass 
deposition rates from the cooling flows but are likely to 
have provided reasonable estimates of the mean cluster 
temperatures.} However, a fit to the central 200 kpc 
radius region (the maximum possible size of any cooling flow; Section 6) 
gives ${\dot M} = 280\pm110$ \Msunpyr~($kT= 8.6^{+1.3}_{-0.9}$ keV), 
in good agreement with the value listed in Table 3.   
These results, and the consistent findings from the spectral, 
image deprojection and X-ray colour profile analyses of the innermost 
100kpc presented in Section 6, highlight the need for 
detailed spatially-resolved spectroscopy when attempting to study 
the properties of cooling flows in distant clusters.

Finally, we note that the excess column density acting on the cooling flow 
component measured with spectral model C1(iii) of $\Delta N_{\rm H} = 
2.4^{+1.2}_{-0.9} \times 10^{21}$\apc~is consistent with the previous 
measurement of $2.9^{+7.6}_{-1.5} \times 10^{21}$\apc~from ASCA, 
using the same model (Allen 2000). 

\section{Measurement of the cluster mass profile}

\subsection{The mass model}

\begin{figure}
\vspace{2cm}
\hbox{
\hspace{0.0cm}\psfig{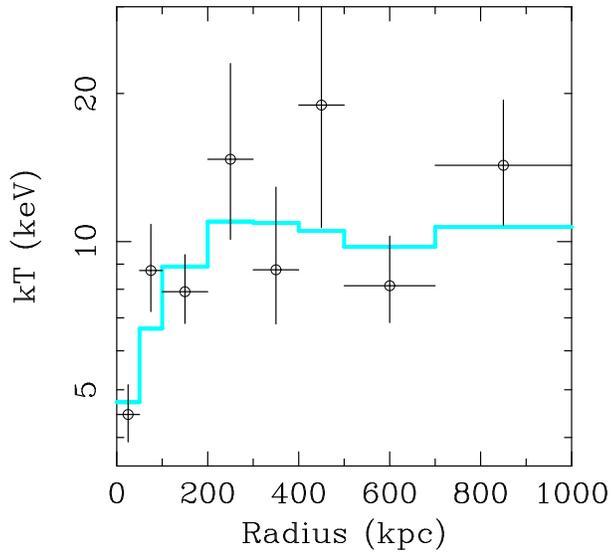}
}
\caption{The predicted deprojected temperature profile (grey curve) determined 
from 100 Monte-Carlo simulations using the best-fitting NFW mass model 
(with $r_{\rm s} = 0.88$ Mpc, $c=4.1$ and $\sigma = 1275$\kmps; see 
Section 5). The predicted profile has been binned to the same spatial 
resolution as the spectral deprojection results (solid points; Section 4.3) 
which are also shown overlaid. The agreement between the deprojected spectral 
results and best-fit NFW mass model predictions (reduced $\chi^2_{\nu} = 1.0$ 
for $\nu=6$ degrees of freedom) indicates that the NFW mass model provides a 
good description of the spatially-resolved Chandra spectra.}\label{fig:kt_xspec3d}
\end{figure}

The observed X-ray surface brightness profile (Fig. \ref{fig:surbri}a)
and deprojected temperature profile (Fig. \ref{fig:kt_xspec3d}) may 
together be used to determine the X-ray gas mass and total mass profiles 
in the cluster. For this analysis, we have used an updated version 
of the image deprojection code developed in Cambridge (see \eg White, 
Jones \& Forman 1997 for details). A variety of simple parameterizations 
for the cluster mass distribution were examined, to establish which 
could provide an adequate description of the Chandra data. For those mass 
models providing reasonable fits, the best-fit parameter values were 
determined using a simple iterative technique.\footnote{Given the 
observed surface brightness profile and a particular parameterized mass 
model, the deprojection code is used to predict the temperature profile of 
the X-ray gas. (We use the median model temperature profile determined from 
100 Monte-Carlo simulations.) The predicted temperature profile is 
rebinned, using an appropriate flux weighting, and compared with the 
results from the spectral deprojection analysis (Fig. \ref{fig:kt_xspec3d}). 
The $\chi^2$ difference between the observed and predicted temperature 
profiles is then calculated. The parameters for the mass model are stepped 
through a regular grid of values in $r_{\rm s}$ and $\sigma$ (or 
alternatively $r_{\rm s}$ and $c$) to determine the best-fit values 
(which give the minimum $\chi^2$) and confidence limits. For Abell 2390 
the final grid used ranges from $0.1-2.5$ Mpc in $r_{\rm s}$ and 
$500-2500$\kmps~in $\sigma$.} Spherical symmetry and hydrostatic 
equilibrium are assumed throughout. 

We find than a good fit ($\chi^2_{\rm min} = 6.1$ for 6 degrees of freedom; 
hereafter DOF) to the Chandra data can be obtained using an NFW mass model

\begin{equation}
\rho(r) = {{\rho_{\rm crit} \delta_{\rm c}} \over {  ({r/r_{\rm s}}) 
\left(1+{r/r_{\rm s}} \right)^2}},
\end{equation}

\noindent where $\rho(r)$ is the mass density, $\rho_{\rm crit} = 3H^2/ 8 \pi G$ is the critical density for closure and 

\begin{equation}
\delta_{\rm c} = {200 \over 3} { c^3 \over \left[ {{\rm ln}(1+c)-{c/(1+c)}}\right]},
\end{equation}

\noindent with a scale radius, $r_{\rm s} = 0.88^{+1.29}_{-0.48}$ Mpc and 
a concentration parameter, $c=4.1^{+2.5}_{-1.7}$ (68 per cent confidence 
limits). The normalization of the mass profile may also be expressed in terms 
of an equivalent velocity dispersion, $\sigma = \sqrt{50} H_0 r_{\rm s} c$ 
(with $r_{\rm s}$ in units of Mpc). The equivalent velocity dispersion 
associated with the best-fit X-ray mass model of 
$\sigma = 1275^{+575}_{-340}$\kmps~(68 per cent confidence limit), 
is in good agreement with the robust, optically-determined value of 
$1262^{+89}_{-68}$\kmps~(Borgani \etal 1999; from a re-analysis of the 
data of Carlberg \etal 1996 data). 

We note that the best-fit values of $c$, $\sigma$ and $r_{\rm s}$ are 
correlated. Fixing $r_{\rm s} = 0.88$ Mpc in the NFW model, we determine 
a concentration parameter $c=4.10\pm0.18$ and an equivalent 
velocity dispersion $\sigma = 1275\pm55$\kmps (68 per cent confidence limit 
keeping $r_{\rm s}$ fixed.) For $r_{\rm s} = 0.4$ Mpc, we obtain 
$c=6.61\pm0.28$ and $\sigma = 935\pm40$\kmps. Fixing $r_{\rm s} = 2.0$ 
Mpc, we obtain  $c=2.53\pm0.11$ and $\sigma = 1790\pm80$\kmps. Note, 
however, that the mass distributions within the central 1Mpc radius are 
similar in all three cases. 

The best-fit NFW mass model, with $r_{\rm s} = 0.88^{+1.29}_{-0.48}$ Mpc 
and $c=4.1^{+2.5}_{-1.7}$, has a virial radius $r_{\rm 200} = c r_{\rm s} = 
3.6^{+1.6}_{-1.0}$ Mpc and an integrated mass within this radius, 
$M_{\rm 200} = 2.7^{+5.6}_{-1.6} \times 10^{15}$\Msun. The relationship 
between $M_{\rm 200}$, $r_{\rm s}$ and $c$ for Abell 2390 is consistent 
with that expected for such a massive cluster formed at a 
redshift $z \sim 0.5-1.0$ (\eg Eke, Navarro \& Frenk 1998)
 
The deprojected X-ray gas temperature profile implied by the best-fitting 
NFW mass model (given the observed surface brightness profile) is shown 
overlaid on the deprojected spectral results in Fig. \ref{fig:kt_xspec3d}. 
The model results have been rebinned to the same resolution as the spectral 
data.

Finally, we note that the NFW mass model provides a significantly better-fit 
to the Chandra data for Abell 2390 than a singular isothermal sphere
($\rho(r) \propto r^{-2}$: $\chi^2_{\rm min} = 33.1$ for 7 DOF). 
However, a variety of other, two-parameter models including a 
softened isothermal sphere ($\rho(r) \propto (1+(r/r_{\rm c})^2)$: 
$\chi^2_{\rm min} = 5.8$ for 6 DOF with $r_{\rm c} \sim 75$ kpc), a 
King approximation to an isothermal sphere ($\rho(r) \propto 
(1+(r/r_{\rm c})^2)^{-3/2}$: $\chi^2_{\rm min} = 6.2$ 
for 6 DOF with $r_{\rm c} \sim 150$ kpc), and a full isothermal sphere 
(Equation 4-125 of Binney \& Tremaine 1987: $\chi^2_{\rm min} = 5.9$ for 
6 DOF with $r_{\rm c} \sim 150$ kpc ) 
also provide good descriptions of the Abell 2390 mass profile.

\subsection{Comparison of X-ray and lensing mass measurements}

Abell 2390 is one of the best studied lensing clusters 
(\eg Pell\'o \etal 1991; Kassiola, Kovner \& Blandford
1992; Narasimha \& Chitre 1993; Pierre \etal 1996; Squires \etal 1996; 
Bezecourt \& Soucail 1997; Frye \& Broadhurst 1998; Pello \etal 1999). 
The cluster exhibits an unusual, strongly lensed `straight arc'
approximately 38 arcsec (174 kpc) away from the nucleus of the 
central galaxy (Pell\'o \etal 1991) in addition to many other 
arcs and arclets (\eg Bezecourt \& Soucail 1997; Pell\'o \etal 1999). 
Pierre \etal (1996) present a two-component mass model for the 
central regions of the cluster and measure a projected
mass within the radius defined by the brightest arc of $1.6 \pm 0.2 
\times 10^{14}$ \Msun. Squires \etal (1996) present a
weak lensing analysis of the cluster and determine an 
azimuthally averaged mass 
profile covering the central $\sim 4.3$ arcmin (1.2 Mpc). 

\begin{figure*}
\vspace{2cm}
\hbox{
\hspace{1.0cm}\psfig{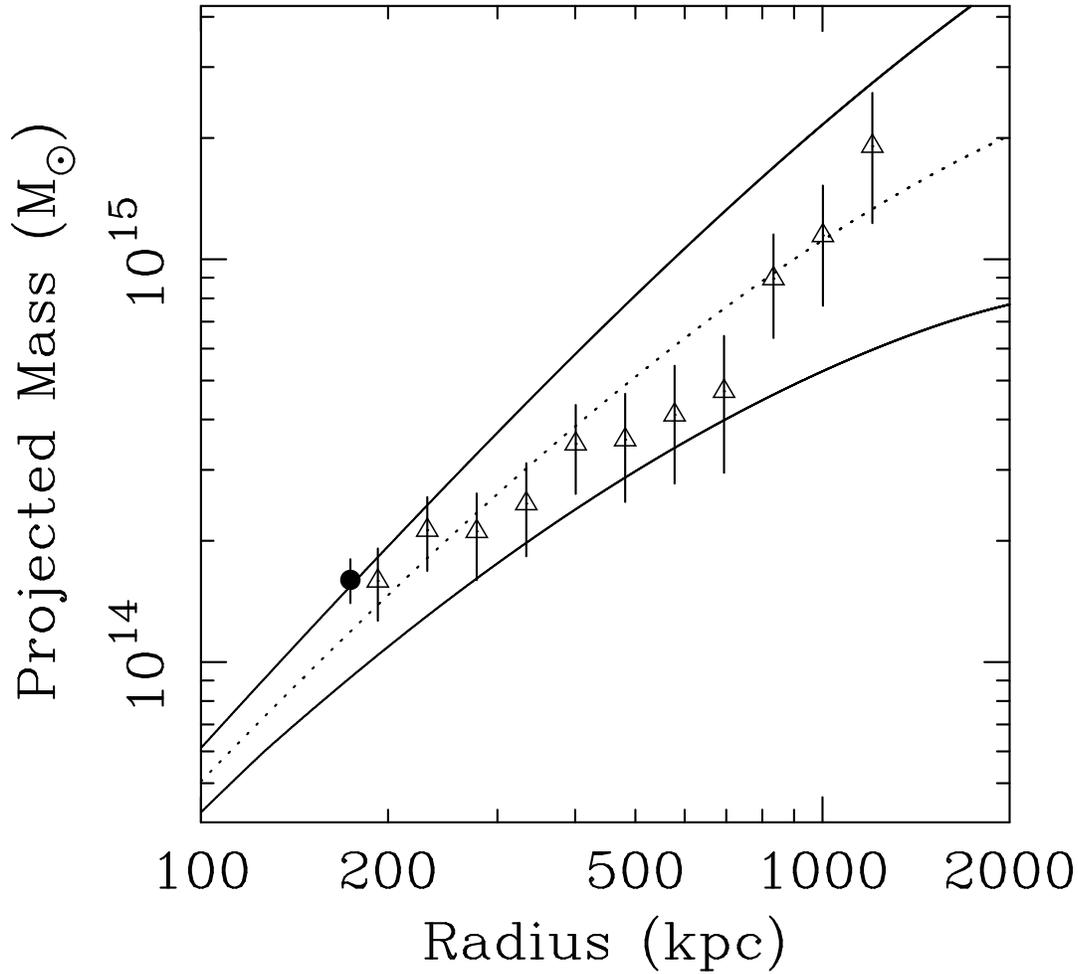}}
\caption{A comparison of the projected total mass determined from the
Chandra X-ray data (Section 5) with the strong lensing result of 
Pierre \etal (1996; filled circle) and the weak lensing results  
of Squires \etal (1996; open triangles). The best-fit NFW 
X-ray mass model is shown as the dotted curve (parameters 
$r_{\rm s} = 880$ kpc, $\sigma = 1275$\kmps). The limits on the 
X-ray results (the maximum and minimum masses at each radius for the 
range of NFW models with $\chi^2 < \chi^2_{\rm min} + 2.30$; the 68 per cent 
confidence contour in the $r_{\rm s}-\sigma$ plane) are shown as solid 
curves.}
\label{fig:xray_lensing}
\end{figure*}

Fig. \ref{fig:xray_lensing} shows the projected mass profile determined 
from the Chandra X-ray data, with the Squires \etal (1996) weak lensing 
results and Piere \etal (1996) strong lensing results overlaid. For the X-ray 
analysis, we assume that the NFW mass models extend out to $r_{\rm 200}$ in 
each case. (The limits on the X-ray results are the maximum and minimum masses 
at each radius for the range of NFW models with $\chi^2 < \chi^2_{\rm min} 
+ 2.30$; the 68 per cent confidence contour in the $r_{\rm s}-\sigma$ plane). 
The agreement between the X-ray and lensing mass results in Fig. 
\ref{fig:xray_lensing} is reasonable at all radii studied. 
The mean scatter of the lensing results about the best-fit X-ray mass 
profile within the central 1 Mpc region is $<20$ per cent. The agreement 
between the independent lensing and X-ray mass measurements, together with 
the consistent results on the equivalent X-ray and observed optical galaxy 
velocity dispersions, confirms the validity of the hydrostatic assumption 
used in the X-ray analysis and suggests that the mass profile in 
Abell 2390 has been robustly determined.

We note that at small radii, the strong lensing mass of Pierre \etal (1996) 
slightly exceeds (at $\sim 1\sigma$ significance) the best-fit value 
determined from the Chandra data ($1.19^{+0.35}_{-0.28} \times 10^{14}$ 
\Msun~within $r=174$ kpc). This difference, albeit marginal, may be related 
to the residual substructure on these scales seen in the X-ray 
(Section 3.1) and optical images and strong lensing mass map 
(Pierre \etal 1996), and could indicate a slight enhancement of the 
central lensing mass due to an alignment of the dominant mass clumps 
in the cluster and/or the presence of additional, non-thermal 
pressure support of the X-ray gas in the cluster core. 

\subsection{The X-ray gas mass fraction}

\begin{figure}
\vspace{2cm}
\hbox{
\hspace{0.0cm}\psfig{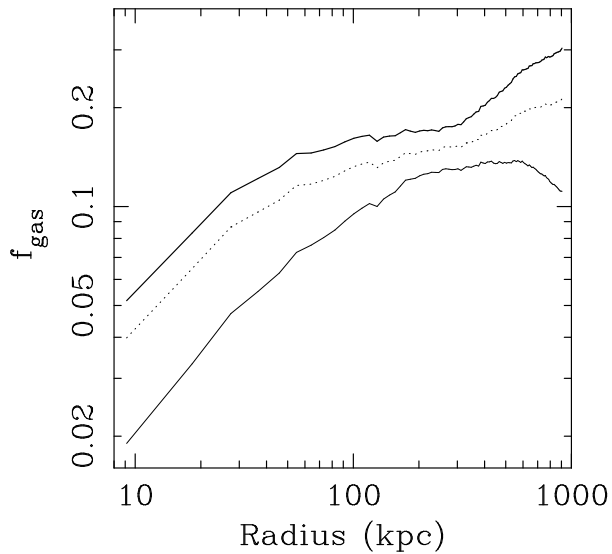}}
\caption{The ratio of the X-ray gas mass to total gravitating mass as 
a function of radius. The three curves show the best-fit value 
(dotted curve) and conservative confidence limits (solid curves;  
determined by combining the $1\sigma$ errors on the integrated gas mass 
at each radius with the uncertainties in the total mass distribution 
shown in Fig. \ref{fig:xray_lensing}). At $r = 0.9$ Mpc 
we measure $f_{\rm gas}=0.21^{+0.09}_{-0.10}$.}\label{fig:baryon}
\end{figure}

The X-ray gas-to-total-mass ratio as a function of radius, $f_{\rm gas}(r)$, 
determined from the Chandra data is shown in Fig. \ref{fig:baryon}. We find 
that the best-fit $f_{\rm gas}$ value rises with increasing radius with 
$f_{\rm gas}=0.21^{+0.09}_{-0.10} h_{\rm 50}^{-1.5}$ at $r=0.9$ Mpc 
(conservative limits determined by combining the $1\sigma$ 
errors on the integrated gas mass at each radius with the uncertainties 
in the total mass distribution shown in Fig. \ref{fig:xray_lensing} in 
quadrature). This value is consistent with the previous 
measurement of $17.5 \pm 2.4$ per cent ($1\sigma$ limits) 
at $r=1$Mpc from Ettori \& Fabian (1999) using ROSAT PSPC and ASCA data 
and assuming strict isothermality in the X-ray gas. 

Following the usual arguments, which assume that the properties of 
clusters provide a fair sample of those of the Universe as a whole 
(\eg White \etal 1993; White \& Fabian 1995; Evrard 1997; 
Ettori \& Fabian 1999; Bahcall \etal 1999), we may use our result 
on the X-ray gas mass fraction in Abell 2390
to estimate the total matter density in the Universe, $\Omega_{\rm m}$. 
Assuming that the luminous baryonic mass in galaxies in Abell 2390 
is approximately one fifth of the X-ray gas mass (\eg White \etal 1993; 
Fukugita, Hogan \& Peebles 1998) and neglecting other possible sources of
baryonic dark matter in the cluster, we obtain $\Omega_{\rm m} =  
(\Omega_{\rm b}/1.2 f_{\rm gas})$, where 
$\Omega_{\rm b}$ is mean baryon density in the Universe and $h_{50}$ is the 
Hubble constant in units of 50\kmpspMpc. For $\Omega_{\rm b}h_{\rm 50}^{2} = 
0.0820\pm0.0072$ (O'Meara \etal 2000),  
$\Omega_{\rm m} = 0.33 \pm 0.16 h_{\rm 50}^{-0.5}$.\footnote{Accounting 
for any additional, dark baryonic matter in the cluster would lower 
the measured value of $\Omega_{\rm m}$. Likewise, if the value of 
$f_{\rm gas}$ increases towards larger radii in Abell 2390, the true value 
for $\Omega_{\rm m}$ will be lower than our quoted result.}

\section{The properties of the cooling flow}

\subsection{Radial properties of the cluster gas}

The results on the electron density and 
cooling time as a function of radius, determined 
from the image deprojection analysis using the best-fit NFW 
mass model, are shown in Fig. \ref{fig:deproj}. Within the central 
500 kpc radius, the electron density profile can be parameterized 
($\chi^2 = 29.1$ for 52 degrees of freedom) by a $\beta$-profile with a 
core radius, $r_c = 21\pm3$ kpc, $\beta=0.378\pm0.014$ 
and a central density, $n_{\rm e}(0)=8.5\pm0.8 \times10^{-2}$ 
cm$^{-3}$ ($1\sigma$ errors). The core radius for the electron 
density distribution is slightly smaller than the value measured directly 
from the projected surface brightness profile under the assumption of strict 
isothermality in the X-ray gas (Section 3.2). Indeed, the evidence 
for any flat central core in the electron density distribution is 
marginal and a simple broken power-law model, with $r_{\rm break} = 
46^{+6}_{-11}$ kpc and slopes interior and exterior to the break radius 
of $-0.48\pm0.05$ and $-1.13\pm0.04$, provides as good a 
fit to the electron density profile in the central 500 kpc region
($\chi^2 = 30.4$ for 51 degrees of freedom).

For an assumed Galactic column density of $6.8\times10^{20}$\apc, we 
measure a central cooling time ({\it i.e.} the mean cooling time within 
the central 1.97 arcsec or ($\sim 9$ kpc) bin of 
$t_{\rm cool}=3.9^{+1.4}_{-0.9} \times 10^8$ yr, and a cooling radius,
at which the cooling time first exceeds 
a Hubble time, of $r_{\rm cool}=175^{+40}_{-6}$ kpc. (Errors on the 
central cooling time are the 10 and 90 percentile values from 100 
Monte Carlo simulations. The upper and lower confidence limits on the 
cooling radii are the points where the 10 and 90 percentile values 
exceed and become less than the Hubble time, respectively.)

\subsection{X-ray colour profile analysis}

\begin{figure}
\vspace{2cm} 
\hbox{
\hspace{0.0cm}\psfig{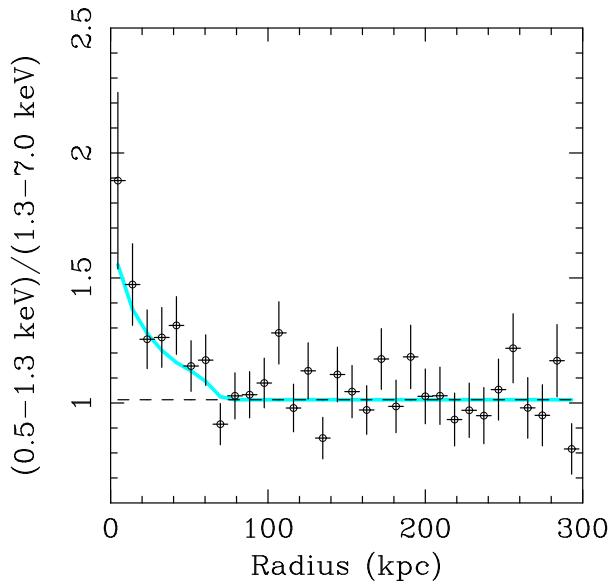}}
\caption{The X-ray colour profile, formed from the ratio of the counts in the 
$0.5-1.3$ and $1.3-7.0$ keV bands, as a function of radius. 
At large radii the ratio is approximately constant and consistent 
with a cluster temperature of $11.0\pm1.5$ keV.
Within a `break' radius of $70^{+20}_{-12}$  kpc, however, the colour ratio 
rises sharply.  The grey curve shows the predicted colour profile 
for a cooling flow of age $\sim 3$ Gyr from the image deprojection 
analysis (Section 6). We use the 
best-fit NFW mass model and assume that the cooling gas is intrinsically 
absorbed by a column density of $2.2 \times 10^{21}$\apc, as determined 
with spectral model C1(iii).}\label{fig:colour}
\end{figure}

We have constructed an X-ray `colour' profile 
for the cluster in order to determine the size 
of the central region in which significant 
cooling occurs. Two separate images were created in 
the energy bands $0.5-1.3$ and $1.3-7.0$ keV 
($0.62-1.60$ and $1.6-8.6$ keV in the rest frame of the source) 
with a 1.97 arcsec (4 raw detector pixels) pixel scale. 
These soft and hard X-ray images were 
background subtracted and flat fielded 
(taking full account of the 
spectral energy distributions of the detected photons). All significant 
point sources were masked out and excluded 
from the analysis. Azimuthally-averaged surface-brightness profiles 
for the cluster were then constructed
in each energy band, centered on the overall peak of the 
X-ray emission (Section 3). The X-ray `colour' profile 
formed from the ratio of the surface brightness profiles in
the soft and hard bands is shown in Fig. \ref{fig:colour}. 

From examination of Fig. \ref{fig:colour} we see that at large radii
the observed X-ray colour ratio is approximately constant, with a mean 
value of $1.01\pm0.03$ ($1\sigma$ error determined from a 
fit to the data between radii of $150-300$ kpc). By comparison with 
simulated spectra we find that this result is consistent 
with an isothermal plasma with a temperature $kT = 11.0\pm1.5$ keV. (We assume 
a metallicity $Z \sim 0.4Z_{\sun}$ and Galactic absorption.) Within a 
`break' radius of $70^{+20}_{-12}$ kpc ($1\sigma$ errors determined from 
a $\chi^2$ fit with a broken power-law model), however, the colour ratio 
rises sharply, indicating the presence of significantly cooler gas.

\subsection{Analysis of the mass deposition profile}

\begin{figure}
\vspace{2cm} 
\hbox{
\hspace{0.0cm}\psfig{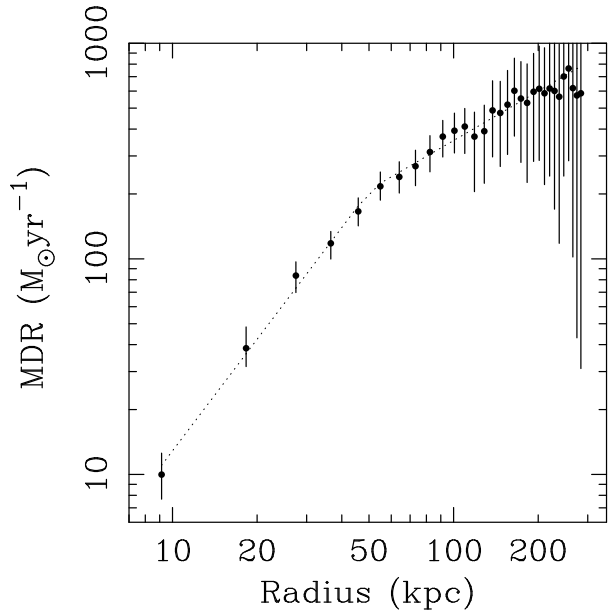}}
\caption{The mass deposition rate (MDR) in units of \Msunpyr, determined from 
the image deprojection analysis using the best-fit NFW mass model. The dotted 
line is the best-fitting
broken power-law model (Section 6.3). Error bars are the 10 and 90 percentile 
values from 1000 Monte Carlo simulations. A Galactic column density of 
$6.8 \times 10^{20}$ \apc, a metallicity of 0.4 solar are assumed.}\label{fig:mdot} 
\end{figure}

The outermost radius at which cooling occurs may also be expected to be 
associated with a `break' in the X-ray surface brightness profile and, 
more evidently, the mass deposition profile determined from the 
deprojection code. The mass deposition profile from the cooling flow, 
which is a parameterization of the X-ray luminosity 
distribution in the cluster core (see \eg White \etal 1997), 
is shown in Fig. \ref{fig:mdot}. Fitting this profile with 
simple a broken power-law model, we determine a break 
radius of $51\pm12$ kpc ($1\sigma$ errors), in reasonable 
agreement with the break radius determined from the fit to the 
X-ray colour profile (Section 6.2) and the electron density 
distribution (Section 6.1) . 
The best-fit broken power-law model is shown
overlaid on the mass deposition profile in Fig. \ref{fig:mdot}. 

The slopes of the mass deposition profile, internal and
external to the break radius are $1.72\pm0.16$ and 
$0.76\pm0.16$, respectively. Accounting only for absorption due to 
cold gas with the nominal Galactic column density, we determine an 
integrated mass deposition rate within the break radius of 
$212^{+37}_{-78}$\Msunpyr. If we also account for the presence of 
intrinsic absorption, with the properties determined using spectral 
model C2(iii), the mass deposition rate within the break radius rises to 
$247^{+43}_{-91}$\Msunpyr, in good agreement with the 
spectral result for the central 100 kpc of $242^{+61}_{-56}$\Msunpyr. 


\subsection{The age of the cooling flow}

Allen \etal (2000a) discuss a number of methods which may be used 
to estimate the ages of cooling flows from X-ray data.  
(Such ages are likely to relate to the time 
intervals since the central regions of clusters were last 
disrupted by major subcluster merger events.) Essentially, 
these methods identify the age of a cooling flow with the 
cooling time of the X-ray gas at the break radius in either the
X-ray colour or deprojected mass deposition profile.

\begin{figure*}
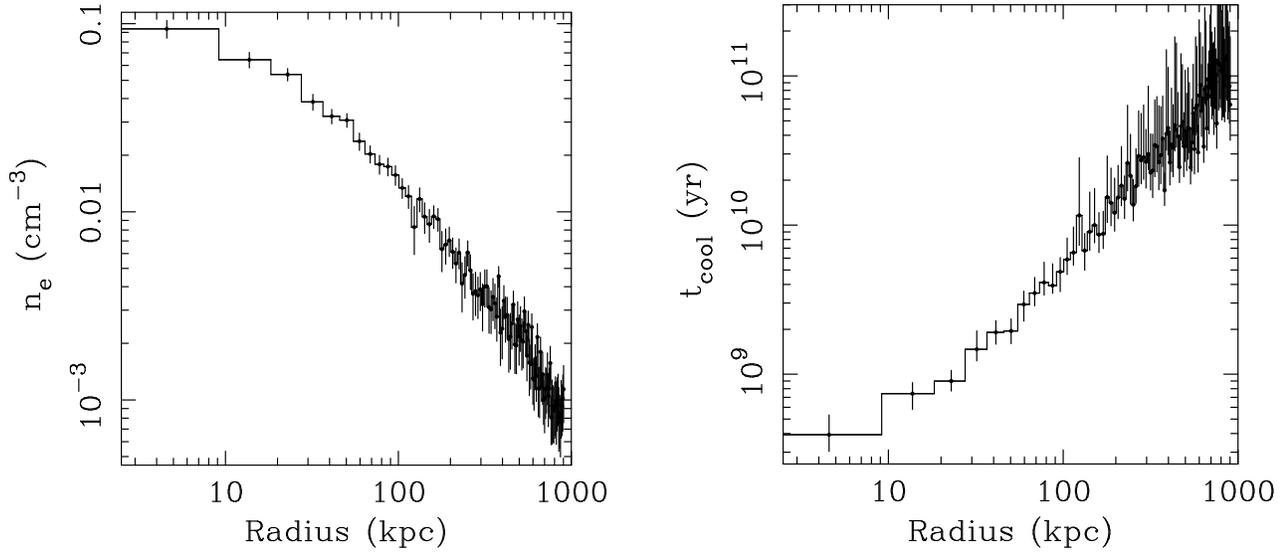

\vspace{2cm}
\hbox{
\hspace{0.2cm}\psfig{figure=ne_deproj_rs088_sigma1275.ps,width=0.45\textwidth,angle=270}
\hspace{0.8cm}\psfig{figure=tcool_deproj_rs088_sigma1275.ps,width=0.45\textwidth,angle=270}
}
\vspace{0.3cm}
\caption{The results on (a) the electron density and (b) the cooling time, 
determined from the X-ray image deprojection analysis using the best-fit NFW mass model. Error bars are the 
$1\sigma$ errors determined from 1000 Monte Carlo simulations. A Galactic 
column density of $6.8 \times 10^{20}$ \apc~and a metallicity of 0.4 solar 
are assumed.}\label{fig:deproj} 
\end{figure*}

Using Fig. \ref{fig:deproj}(b), we see that cooling time of the 
X-ray gas at the break radius in the X-ray colour profile 
($70^{+20}_{-12}$ kpc) lies in the range $3.2^{+0.9}_{-0.6}$ Gyr. (The 
cooling time at the break radius is measured from a least-squares fit 
to the data in Fig. \ref{fig:deproj}(b) over the $10-100$ kpc range using 
a power-law model.) If we instead identify the age of the 
cooling flow with the cooling time at the break radius in the mass 
deposition profile in Fig. \ref{fig:mdot} ($51\pm12$ kpc), 
we infer an age for the cooling flow of $2.3^{+0.6}_{-0.5}$ Gyr. 
(In both cases we assume that no intrinsic 
absorption acts beyond the outer edge of the cooling flow, which is 
reasonable if the absorbing matter is accumulated by the flow.)

In summary, we see the X-ray colour profile, image deprojection 
analysis and spectral data provide consistent results on 
the properties of the cooling flow in Abell 2390, indicating a
mass deposition rate in the range  $200-300$ \Msunpyr~and an age 
of $2-3$ Gyr. 

\subsection{On the rising ambient temperature profile within the 
cluster core}

In principle, the results on the X-ray gas temperature and 
surface brightness within the cluster core may be used to 
distinguish between an NFW model for the dark matter distribution and 
alternative models with a steeper central cusp (\eg Moore \etal 1999). 
The results for Abell 2390 presented here are certainly consistent with 
an NFW profile (see Fig. \ref{fig:kt_xspec3d}). However, given the 
complexity of the gas within the central 100 kpc and the relatively 
short exposure time of the present observation (which limits 
the number of independent spectra which can be extracted from the region 
of the cluster core), we do not attempt to explore such models here. 
This issue will be better addressed with Chandra data for nearer, brighter 
systems. 

The fact that the ambient temperature profile, corrected for the 
effects of the cooling flow, rises with increasing radius 
throughout the central $\sim 200$ kpc, is interesting. In detail, 
the density and temperature structure in a cluster core will depend on 
the thermal history of the gas, as well as the underlying dark matter 
distribution and, 
subject to pressure equilibrium and convective stability, may be flat, 
increase or decrease with radius. As discussed in Section 4.6, the 
presence of relatively cool, ambient gas in the central regions of  
Abell 2390 may have lead to overestimates of the mass deposition rate 
in previous studies with ASCA and Beppo-SAX in which this region was 
not spatially resolved.

There are several reasons why relatively cool, dense gas might 
exist beyond the outer edge of the present-day cooling 
flow in Abell 2390. The first 
is that a pre-existing cooling flow may have been disturbed several Gyr ago. 
Cooler, denser gas could then have been spread out over several 100 kpc, 
with the cooling flow having so far only re-established itself within the 
inner $\sim 50-70$ kpc. A second possibility is that the densest gas from the 
cores of infalling subclusters may have been stripped and deposited over 
time in the core of the main cluster without strong shocking (Fabian \& 
Daines 1991). The situation seen in the Chandra data for Abell 2142 
(Markevitch \etal 2000) is reminiscent of this, with a sharp drop in 
temperature from $kT \sim 14$ keV to $kT \sim 7$ keV (and a corresponding 
rise in density) observed at $r \sim 370$ kpc, moving inwards from 
northwest of the cluster centre. Such a sharp drop in temperature at these 
radii cannot be due to any current cooling flow in the cluster.

\section{Conclusions}

The main conclusions from this work may be summarized as follows:
\vskip 0.2cm

(i) We have measured the distribution of mass in Abell 2390 using Chandra 
X-ray observations. The mass profile can be well-modelled by an NFW profile 
with a scale radius $r_{\rm s} = 0.88^{+1.29}_{-0.48}$ Mpc and a 
concentration parameter, $c=4.1^{+2.5}_{-1.7}$ (68 per cent confidence 
limits). The normalization of the mass profile may also be expressed 
in terms of an equivalent velocity dispersion, $\sigma = \sqrt{50} 
H_0 r_{\rm s} c = 1275^{+575}_{-340}$\kmps, in good agreement with the 
optically-determined value of $1262^{+89}_{-68}$\kmps~(Borgani 
\etal 1999).

(ii) The best-fit Chandra mass model is in good agreement 
with independent measurements from strong and weak lensing studies. The
mean scatter between the X-ray and lensing values within the central 
$r = 1$ Mpc radius is $<20$ per cent. 

(iii) The X-ray gas to total mass ratio rises with increasing radius 
within the central 0.9 Mpc, with $f_{\rm gas}=0.21^{+0.09}_{-0.10}$ 
at $r=0.9$ Mpc. Following the usual arguments, this result 
on the X-ray gas mass fraction may be converted into a constraint on the 
mean mass density in the Universe, $\Omega_{\rm m} = 
0.33 \pm 0.16 h_{\rm 50}^{-0.5}$.

(iv) The X-ray gas temperature rises with increasing radius within the 
central $r \sim 200$ kpc, and then remains approximately isothermal with 
$kT = 11.5^{+1.6}_{-1.3}$ keV out to $r \sim 1.0$ Mpc. 

(v) The azimuthally-averaged $0.3-7.0$ keV surface brightness profile 
exhibits a small core radius and a clear break at $r \sim 0.5$ Mpc, where the 
slope changes abruptly from $S_{\rm X} \approxpropto r^{-1.5}$ to 
$S_{\rm X} \approxpropto r^{-3.6}$. 

(vi) The best-fit mass deposition rate from the cooling flow, determined in a 
consistent manner from the spectral and imaging Chandra data, lies in the 
range $200-300$ \Msunpyr. This value is lower than previous estimates 
based on integrated ASCA and Beppo-SAX spectra for the entire cluster, 
which could not resolve the drop in the central, ambient gas 
temperature. We estimate an age for the cooling flow of $2-3$ Gyr.

\section*{Acknowledgements}

We thank G. Squires for communicating his weak lensing results, R. 
Mushotzky for helpful comments, and P. Thomas for helpful comments 
and discussions. We also thank R. Johnstone and R. Schmidt for coding and 
discussions regarding the analysis of Chandra data. We acknowledge the 
support of the Royal Society.

\clearpage

\begin{table*}
\begin{center}
\caption{The results from the analysis of the annular spectra 
using spectral model A in the $0.5-7.0$ keV band. Temperatures ($kT$) are in
keV, metallicities ($Z$) in solar units and normalizations ($K$) in units of 
$(10^{-17}/(4 \pi (D_{\rm A}(1+z))^2)) \int n_{\rm e} n_{\rm H} dV$, 
where $D_{\rm A}$ is the angular size distance to the source in cm, 
$n_{\rm e}$ is the electron density (in cm$^{-3}$), and $n_{\rm H}$ 
is the hydrogen ion density (in cm$^{-3}$). The absorbing column density has been 
fixed at the nominal Galactic value of $6.8 \times 10^{20}$\apc. The total $\chi^2$ values 
and number of degrees of freedom (DOF) in the fits are listed in column 5. 
Error bars at both the $1\sigma$ ($\Delta \chi^2=1.0$) and 90 per cent
($\Delta \chi^2=2.71$; in parentheses) confidence limits on a single interesting parameter are listed. 
For the outermost $0.7-1.0$ Mpc annulus, where the instrumental background is 
most significant, only data in the energy range $0.5-6.0$ keV range were used.}
\vskip 0.2truein
\begin{tabular}{ c c c c c c }
&&&&&  \\                                                                                                                   \hline
Model A    & ~ &  $kT$                    &    $Z$                 &      $K$                  &        $\chi^2$/DOF \\
								                               
\hline                                                                                                                                              
$0-50$     & ~ &  $5.58^{+0.52(+0.93)}_{-0.42(-0.65)}$  & $0.48^{+0.16(+0.29)}_{-0.18(-0.29)}$ & $1.48^{+0.07(+0.11)}_{-0.06(-0.10)}$ &98.2/79   \\
$0-100$    & ~ &  $6.81^{+0.43(+0.74)}_{-0.39(-0.62)}$  & $0.48^{+0.11(+0.18)}_{-0.10(-0.17)}$ & $3.41^{+0.08(+0.14)}_{-0.09(-0.14)}$ &132.5/123   \\
$0-200$    & ~ &  $7.97^{+0.45(+0.77)}_{-0.42(-0.67)}$  & $0.36^{+0.08(+0.13)}_{-0.08(-0.13)}$ & $6.49^{+0.11(+0.19)}_{-0.12(-0.19)}$ &159.7/163   \\
$50-100$   & ~ &  $8.58^{+1.07(+1.86)}_{-0.85(-1.31)}$  & $0.55^{+0.19(+0.33)}_{-0.19(-0.31)}$ & $1.86^{+0.07(+0.11)}_{-0.07(-0.12)}$ &96.9/98   \\
$100-150$  & ~ &  $9.54^{+1.45(+2.60)}_{-1.10(-1.72)}$  & $0.21^{+0.17(+0.29)}_{-0.17(-0.21)}$ & $1.62^{+0.07(+0.10)}_{-0.06(-0.09)}$ &78.9/82   \\
$150-200$  & ~ &  $8.94^{+1.40(+2.64)}_{-1.06(-1.62)}$  & $0.48^{+0.28(+0.47)}_{-0.27(-0.45)}$ & $1.32^{+0.07(+0.12)}_{-0.11(-0.07)}$ &69.0/74   \\
$100-200$  & ~ &  $9.40^{+0.97(+1.75)}_{-0.82(-1.28)}$  & $0.28^{+0.14(+0.24)}_{-0.14(-0.23)}$ & $2.98^{+0.09(+0.15)}_{-0.09(-0.14)}$ &128.9/121   \\
$200-300$  & ~ & $12.03^{+1.93(+3.45)}_{-1.60(-2.42)}$  & $0.26^{+0.23(+0.38)}_{-0.24(-0.26)}$ & $2.35^{+0.11(+0.16)}_{-0.09(-0.15)}$ &106.7/108   \\
$300-400$  & ~ & $10.84^{+1.76(+3.40)}_{-1.39(-2.13)}$  & $0.37^{+0.25(+0.40)}_{-0.25(-0.37)}$ & $2.07^{+0.11(+0.17)}_{-0.09(-0.14)}$ &100.6/105   \\
$400-500$  & ~ & $12.46^{+2.33(+4.30)}_{-1.77(-2.70)}$  & $0.00^{+0.10(+0.23)}_{-0.00(-0.00)}$ & $1.73^{+0.04(+0.07)}_{-0.04(-0.07)}$ &89.1/88   \\
$500-700$  & ~ &  $9.73^{+1.56(+2.80)}_{-1.19(-1.81)}$  & $0.49^{+0.27(+0.46)}_{-0.26(-0.44)}$ & $2.14^{+0.11(+0.19)}_{-0.10(-0.16)}$ &62.5/67   \\
$700-1000$ & ~ & $13.08^{+4.51(+8.49)}_{-2.81(-3.99)}$  & $0.55^{+0.61(+1.06)}_{-0.55(-0.55)}$ & $1.37^{+0.14(+0.21)}_{-0.11(-0.18)}$ &46.1/52   \\
$200-1000$ & ~ & $11.46^{+0.95(+1.62)}_{-0.79(-1.26)}$  & $0.23^{+0.12(+0.20)}_{-0.13(-0.21)}$ & ---  &410.4/428   \\
&&&&& \\
\hline
\end{tabular}
\end{center}
 \parbox {7in}
{}
\vskip 0.5truein
\end{table*}

\begin{table*}
\begin{center}
\caption{The results from the analysis of the annular spectra 
in the $0.5-7.0$ keV band using spectral model B. The absorbing
column density ($N_{\rm H}$) is in units of $10^{20}$ \apc. 
Other details as in Table 1.}
\vskip 0.2truein
\begin{tabular}{ c c c c c c }
&&&&&  \\                                                                                                                 
\hline                                                                                                                    
Model B    & ~ &  $kT$                    &    $Z$                 & $N_{\rm H}$             & $\chi^2$/DOF \\
\hline                                                                                                                   
$0-50$     & ~ &  $4.77^{+0.47(+0.82)}_{-0.40(-0.63)}$  & $0.48^{+0.16(+0.26)}_{-0.14(-0.23)}$ & $9.86^{+1.25(+2.08)}_{-1.24(-2.00)}$  & 91.8/78   \\
$0-100$    & ~ &  $6.18^{+0.49(+0.83)}_{-0.41(-0.66)}$  & $0.45^{+0.09(+0.16)}_{-0.10(-0.16)}$ & $8.71^{+0.95(+1.58)}_{-0.94(-1.53)}$  & 128.3/122 \\
$0-200$    & ~ &  $7.19^{+0.50(+0.87)}_{-0.44(-0.70)}$  & $0.33^{+0.07(+0.12)}_{-0.07(-0.11)}$ & $8.35^{+0.71(+1.17)}_{-0.70(-1.14)}$  & 154.7/162 \\
$50-100$   & ~ &  $7.46^{+1.06(+1.97)}_{-0.77(-1.21)}$  & $0.49^{+0.18(+0.30)}_{-0.16(-0.26)}$ & $8.83^{+1.31(+2.18)}_{-1.32(-2.16)}$  & 94.5/97\\
$100-150$  & ~ &  $8.31^{+1.52(+2.83)}_{-1.19(-1.84)}$  & $0.22^{+0.15(+0.25)}_{-0.16(-0.22)}$ & $8.48^{+1.42(+2.37)}_{-1.37(-2.22)}$  & 77.3/81\\
$150-200$  & ~ &  $8.32^{+1.68(+3.30)}_{-1.15(-1.74)}$  & $0.47^{+0.26(+0.45)}_{-0.25(-0.41)}$ & $7.78^{+1.58(+2.63)}_{-1.56(-2.53)}$  & 68.6/73\\
$100-200$  & ~ &  $8.62^{+1.14(+2.06)}_{-0.91(-1.40)}$  & $0.28^{+0.13(+0.21)}_{-0.13(-0.21)}$ & $7.90^{+1.06(+1.75)}_{-1.04(-1.69)}$  & 127.8/120\\
$200-300$  & ~ & $12.03^{+2.79(+5.16)}_{-2.20(-3.17)}$  & $0.26^{+0.22(+0.38)}_{-0.24(-0.26)}$ & $6.80^{+1.30(+2.15)}_{-1.18(-1.95)}$  & 106.7/107\\
$300-400$  & ~ &  $8.79^{+1.74(+3.26)}_{-1.20(-1.85)}$  & $0.41^{+0.20(+0.33)}_{-0.20(-0.34)}$ & $8.97^{+1.35(+2.26)}_{-1.35(-2.22)}$  & 98.0/104\\
$400-500$  & ~ & $12.52^{+3.52(+7.23)}_{-2.32(-3.49)}$  & $0.00^{+0.10(+0.23)}_{-0.00(-0.00)}$ & $6.76^{+1.51(+2.50)}_{-1.49(-2.47)}$  & 89.1/87\\
$500-700$  & ~ & $10.23^{+2.49(+4.97)}_{-1.75(-2.55)}$  & $0.48^{+0.29(+0.48)}_{-0.28(-0.43)}$ & $6.34^{+1.34(+2.21)}_{-1.29(-2.11)}$  & 62.3/66\\
$700-1000$ & ~ & $10.92^{+5.05(+11.0)}_{-2.42(-3.54)}$  & $0.52^{+0.52(+0.99)}_{-0.46(-0.52)}$ & $8.23^{+1.99(+3.34)}_{-2.04(-3.37)}$  & 45.5/51\\
$200-1000$ & ~ & $10.87^{+1.19(+2.16)}_{-1.02(-1.59)}$  & $0.25^{+0.11(+0.19)}_{-0.13(-0.21)}$ & $7.28^{+0.66(+1.09)}_{-0.64(-1.06)}$  & 409.9/427   \\
&&&&& \\
\hline
\end{tabular}
\end{center}
 \parbox {7in}
{}
\end{table*}

\begin{table*}
\vskip 0.2truein
\begin{center}
\caption{The best-fit parameter values and $1\sigma$ (and 90 per cent; in parentheses) 
confidence limits determined from the multiphase analysis of the central 100 kpc region in 
the extended $0.3-7.0$ keV band. The mass deposition rates from the cooling flow
(${\dot M}$) are in units of \Msunpyr. Column densities ($N_{\rm H}$) and  
intrinsic column densities ($\Delta N_{\rm H}$) are in units of $10^{20}$ atom 
cm$^{-2}$. The limits on the edge energy ($E_{\rm edge}$) determined with absorption model (v) 
are in keV.  Other details as in Table 1.}
\vskip 0.2truein
\begin{tabular}{ c c c c c c c }
\hline                                                              
\multicolumn{1}{c}{} &
\multicolumn{1}{c}{} &
\multicolumn{1}{c}{} &
\multicolumn{3}{c}{EMISSION MODEL} \\
\multicolumn{1}{c}{} &
\multicolumn{1}{c}{} &
\multicolumn{1}{c}{} &
\multicolumn{1}{c}{SINGLE-PHASE~} &
\multicolumn{2}{c}{MULTIPHASE} \\
ABSORPTION MODEL &       & ~ &        A               &       C1                &            C2             \\
\hline
&&&&&& \\
&$kT_1$                  & ~ & $8.14^{+0.64(+1.14)}_{-0.56(-0.88)}$ & $8.14^{+0.67(+1.36)}_{-0.56(-0.88)}$  & $8.14^{+0.66(+1.43)}_{-0.56(-0.88)}$    \\   
CASE (i)   &$Z$          & ~ & $0.50^{+0.13(+0.21)}_{-0.13(-0.21)}$ & $0.50^{+0.13(+0.21)}_{-0.12(-0.21)}$  & $0.50^{+0.13(+0.21)}_{-0.12(-0.21)}$    \\
GALACTIC ABSORPTION &${\dot M}$   & ~ & ---                         & $0^{+17(+38)}_{-0(-0)}$               & $0^{+17(+38)}_{-0(-0)}$                 \\
&$\chi^2$/DOF            & ~ & 203.1/133                            & 203.1/132                             & 203.1/132                               \\
&&&&&& \\												  
													  
&$kT_1$                  & ~ & $5.78^{+0.39(+0.67)}_{-0.35(-0.57)}$ & $6.59^{+0.78(+1.47)}_{-0.66(-1.00)}$  & $6.45^{+0.73(+1.33)}_{-0.59(-0.90)}$    \\
CASE (ii)  &$Z$          & ~ & $0.40^{+0.10(+0.16)}_{-0.07(-0.14)}$ & $0.41^{+0.10(+0.17)}_{-0.15(-0.09)}$  & $0.42^{+0.09(+0.16)}_{-0.10(-0.16)}$    \\
VARIABLE ABSORPTION 
           &${\dot M}$   & ~ & ---                                  & $109^{+58(+98)}_{-58(-94)}$           & $109^{+60(+100)}_{-57(-93)}$            \\
BY COLD GAS (z=0)   
         &$N_{\rm H}$    & ~ & $10.80^{+0.58(+0.96)}_{-0.56(-0.91)}$& $11.40^{+0.71(+1.19)}_{-0.68(-1.09)}$ & $11.40^{+0.70(+1.19)}_{-0.68(-1.09)}$   \\
&$\chi^2$/DOF            & ~ & 141.6/132                            & 137.9/131                             & 137.9/131                               \\
&&&&&& \\												  
													  
&$kT_1$                  & ~ & $5.90^{+0.39(+0.67)}_{-0.35(-0.57)}$ & $8.39^{+1.29(+2.47)}_{-0.88(-1.33)}$  & $7.39^{+0.93(+1.75)}_{-0.58(-1.04)}$    \\
CASE (iii)           &$Z$& ~ & $0.41^{+0.10(+0.16)}_{-0.09(-0.14)}$ & $0.44^{+0.11(+0.19)}_{-0.11(-0.18)}$  & $0.43^{+0.11(+0.19)}_{-0.10(-0.17)}$    \\
INTRINSIC ABSORPTION
      &${\dot M}$        & ~ &          ---                         & $270^{+40(+68)}_{-37(-61)}$           & $242^{+37(+61)}_{-35(-56)}$             \\
BY COLD GAS  ($z=z_{\rm clus}$)
&$\Delta N_{\rm H}$      & ~ & $5.37^{+0.77(+1.27)}_{-0.75(-1.22)}$ & $24.02^{+6.91(+12.14)}_{-5.96(-9.26)}$& $21.96^{+6.06(+10.75}_{-5.08(-7.92)}$  \\
&$\chi^2$/DOF            & ~ & 143.5/132                            & 143.7/131                             & 142.9/131                 \\
													  
&&&&&& \\												  
&$kT_1$                  & ~ & $5.90^{+0.39(+0.67)}_{-0.35(-0.57)}$ & $8.38^{+1.30(+2.49)}_{-0.87(-1.32)}$  & $7.45^{+0.88(+1.69)}_{-0.73(-1.10)}$    \\
CASE (iv)        &$Z$    & ~ & $0.41^{+0.10(+0.16)}_{-0.09(-0.14)}$ & $0.44^{+0.11(+0.19)}_{-0.11(-0.17)}$  & $0.43^{+0.11(+0.19)}_{-0.10(-0.17)}$    \\
PARTIAL COVERING 
           &${\dot M}$   & ~ & ---                                  & $270^{+40(+65)}_{-37(-61)}$           & $246^{+46(+112)}_{-38(-60)}$        \\
BY COLD GAS 												   
 ($z=z_{\rm clus}$) 
    &$\Delta N_{\rm H}$  & ~ & $5.37^{+0.77(+1.27)}_{-0.75(-1.22)}$ & $24.07^{+6.86(+12.08)}_{-6.01(-9.31)}$& $21.35^{+8.38(+17.93)}_{-4.52(-7.37)}$  \\
&$f$                     & ~ & $1.00^{+0.00(+0.00)}_{-0.29(-0.47)}$ & $1.00^{+0.00(+0.00)}_{-0.16(-0.33)}$  & $1.00^{+0.00(+0.00)}_{-0.17(-0.32)}$    \\
&$\chi^2$/DOF            & ~ & 143.5/131                            & 143.7/130                             & 142.9/130                 \\
&&&&&& \\												  
													  
&$kT_1$                  & ~ & $6.25^{+0.35(+0.71)}_{-0.36(-0.58)}$ & $8.54^{+1.06(+1.93)}_{-0.81(-1.25)}$  & $7.67^{+0.81(+1.50)}_{-0.67(-1.04)}$    \\ 
CASE (v)    &$Z$         & ~ & $0.41^{+0.10(+0.16)}_{-0.09(-0.15)}$ & $0.43^{+0.11(+0.19)}_{-0.12(-0.20)}$  & $0.42^{+0.12(+0.20)}_{-0.11(-0.18)}$    \\
SIMPLE EDGE 
           &${\dot M}$   & ~ & ---                                  & $214^{+32(+55)}_{-29(-47)}$           & $189^{+28(+46)}_{-27(-43)}$         \\
($z=z_{\rm clus}$) 
&$E_{\rm edge}$          & ~ & $0.43(<0.54)$                        & $<0.41(<0.43)$                        & $<0.41(<0.43)$                   \\
&$\tau$                  & ~ & $1.02^{+0.21(+1.85)}_{-0.16(-0.26)}$ & $17.7^{+6.89(+10.5)}_{-11.9(-13.4)}$  & $15.5^{+5.9(+9.9)}_{-10.2(-11.5)}$     \\
&$\chi^2$/DOF            & ~ & 135.8/131                            & 141.9/130                             & 140.6/130                 \\
&&&&& \\ 

\hline    
\end{tabular}
\end{center}
\parbox {7in}
{}
\end{table*}

\clearpage

\clearpage

\clearpage

\clearpage

\clearpage

\clearpage

\clearpage 

\clearpage

\clearpage

\clearpage

\clearpage

\clearpage

\clearpage

\begin{thebibliography}{}

\bibitem{} Akritas M.G., Bershady M.A., 1996, ApJ, 470, 706 
\bibitem{} Allen S.W., 1998, MNRAS, 296, 392
\bibitem{} Allen S.W., 2000, MNRAS, 315, 269
\bibitem{} Allen S.W., Fabian A.C., 1997, MNRAS, 286, 583
\bibitem{} Allen S.W., Fabian A.C., 1998, MNRAS, 297, L57
\bibitem{} Allen S.W., Fabian A.C., Kneib J.-P., 1996, MNRAS, 279, 615
\bibitem{} Allen S.W., Fabian A.C., Johnstone R.M., Nulsen P.E.J., Arnaud K.A., 2000a, MNRAS, submitted
\bibitem{} Allen S.W. \etal 2000b, MNRAS, submitted
\bibitem{} Anders E., Grevesse N., 1989, Geochemica et Cosmochimica Acta 53, 197
\bibitem{} Arnaud, K.A., 1996, in Astronomical Data Analysis Software and Systems V, eds. Jacoby G. and Barnes J., ASP Conf. Series volume 101, p17
\bibitem{} Arnaud K.A., Mushotzky R.F., 1998, ApJ, 501, 119
\bibitem{} Bacon D.J., Refregier A., Clowe D., Ellis R., 2000, MNRAS, 318, 625
\bibitem{} Bahcall N.A., Ostriker J.P., Perlmutter S., Steinhardt P.J., 1999, Science, 284, 1481
\bibitem{} Balucinska-Church M., McCammon D., 1992, ApJ, 400, 699
\bibitem{} Bartelmann M., Steinmetz M., 1996, MNRAS, 283, 431
\bibitem{} Bartelmann M., Schneider P., preprint (astro-ph/9912508)
\bibitem{} Bezecourt J., Soucail G., 1997, A\&A, 317, 661 
\bibitem{} B\"ohringer H., Tanaka Y., Mushotzky R.F., Ikebe Y., Hattori M., 1998, A\&A, 334, 789
\bibitem{} Borgani S., Girardi M., Carlberg R.G., Yee H.K.C., Ellingson E., 1999, ApJ, 527, 561
\bibitem{} Buote D.A., 2000, ApJ, 532, L113
\bibitem{} Buote D.A., Tsai J.C., 1996, ApJ, 458, 27
\bibitem{} Carlberg R.G., Yee H.K.C., Ellingson E., Abraham R., Gravel P., Morris S., Pritchet C.J., 1996, ApJ, 462, 32
\bibitem{} de Grandi S., Molendi S., 1999, ApJ, 527, L25
\bibitem{} den Hartog R., Katgert P., 1996, MNRAS, 279, 349
\bibitem{} Dickey J.M., Lockman F.J., 1990, ARA\&A, 28, 215 
\bibitem{} Ebeling H., Rangarajan F.V.N., White D.A., 2000, MNRAS, in press
\bibitem{} Edge A.C., Stewart G.C., Fabian A.C., 1992, MNRAS, 255, 431
\bibitem{} Edge A.C., Ivison R.J., Smail I., Blain A.W., Kneib J.-P., 1999, MNRAS, 306, 599
\bibitem{} Eke V.R., Navarro J.F., Frenk C.S., 1998, ApJ, 503, 569
\bibitem{} Ettori S., Fabian A.C., 1999, MNRAS, 305, 834
\bibitem{} Ettori S., Allen S.W., Fabian A.C., 2000, MNRAS, in press
\bibitem{} Evrard A.E., 1997, MNRAS, 292, 289
\bibitem{} Fabian A.C., Daines S.J., 1991, MNRAS, 252, L17
\bibitem{} Fabian A.C., Hu E.M., Cowie L.L., Grindlay J., 1981, ApJ, 248, 47
\bibitem{} Fabian A.C. \etal 2000a, MNRAS, 315, L8
\bibitem{} Fabian A.C. \etal 2000b, MNRAS, in press 
\bibitem{} Fadda D., Girardi M., Giuricin G., Mardirossian F., Mezzetti M., 1996, ApJ, 473, 670
\bibitem{} Fort B., Mellier Y., 1994, A\&AR, 5, 239
\bibitem{} Frenk C.S., White S.D.M., Efstathiou G., Davis M., 1990, ApJ, 351, 10
\bibitem{} Frye B., Broadhurst T., 1998, ApJL, 499, 115
\bibitem{} Fukugita M., Hogan C.J., Peebles P.J.E., 1998, ApJ, 503, 518
\bibitem{} Geller M.J., Diaferio A., Kurtz M.J., 1999, ApJ, 517, 23
\bibitem{} Hutchings J.B., Balogh M.L., 2000, AJ, 119, 1123
\bibitem{} Irwin J.A., Bregman J.N., 2000, ApJ, 538, 543
\bibitem{} Johnstone R.M., Fabian A.C., Edge A.C., Thomas P.A., 1992, MNRAS, 255, 431
\bibitem{} Jones C., Forman W., 1984, ApJ, 276, 38
\bibitem{} Kaastra J.S., Mewe R., 1993, Legacy, 3, HEASARC, NASA
\bibitem{} Kassiola A., Kovner I., Blandford R.D., 1992, ApJ, 396, 10
\bibitem{} Kikuchi K., Tae F., Ezawa H., Yamasaki N.Y., Ohashi T., Fukazawa Y., Ikebe Y., 1999, PASJ, 51, 301
\bibitem{} Kneib J.-P., Mellier Y., Pell\'o R., Miralda-Escud\'e J., Le Borgne J.-F., B\"ohringer H., Picat J.-P., 1995, A\&A, 303,27
\bibitem{} Koranyi D.M., Geller M.J., 2000, AJ, 119, 44
\bibitem{} Kriss G.A., Cioffi D.F., Canizares C.R., 1983, ApJ, 272, 439 
\bibitem{} Lemonon L., Pierre M., Cesarsky C.J., Elbaz D., Pell\'o R., Soucail G., Vigroux L., 1998, A\&A, 334, L21
\bibitem{} Lewis A.D., Ellingson E., Morris S.L., Carlberg R.G., 1999, 517, 587
\bibitem{} Liedhal D.A., Osterheld A.L., Goldstein W.H., 1995, ApJ, 438, L115 
\bibitem{} Loeb A., Mao, S., 1994, ApJ, 435, 109L 
\bibitem{} Lucey J.R., 1983, MNRAS, 204, 33 
\bibitem{} Markevitch M., Forman W.R., Sarazin C.L., Vikhlinin A., 1998, ApJ, 503, 77
\bibitem{} Markevitch M. \etal, 2000, ApJ, 541, 542
\bibitem{} Mazure A., \etal, 1996, A\&A, 310, 31
\bibitem{} McNamara B.R. \etal, 2000, ApJL, 534, 135
\bibitem{} Mellier Y., 1999, ARA\&A, 37, 127
\bibitem{} Miralda-Escud\'e J., Babul A., 1995, ApJ, 449, 18
\bibitem{} Moore B., Quinn T., Governato F., Stadel J., Lake G., 1999, MNRAS, 310, 1147
\bibitem{} Narasimha D., Chitre S.M., 1993, A\&A, 280, 57
\bibitem{} Navarro J.F., Frenk C.S., White S.D.M., 1997,ApJ, 490, 493 
\bibitem{} Nulsen P.E.J., 1998, MNRAS, 297, 1109
\bibitem{} O'Meara J.M., Tytler D., Kirkman D., Suzuki N., Prochaska J.X., Lubin D., Wolfe A.M., 2000, ApJ, submitted (astro-ph/0011179)
\bibitem{} Pell\'o R., Sanahuja B., Le Borgne J.F., Soucail G., Mellier Y., 1991, ApJ, 366, 405
\bibitem{} Pell\'o R. \etal, 1999, A\&A, 346, 359
\bibitem{} Pierre M., Le Borgne J.F., Soucail G., Kneib J.-P., 1996, A\&A, 311, 413
\bibitem{} Roettiger K., Burns J.O., Loken C., 1996, ApJ, 473, 651
\bibitem{} Sarazin C.L., 1988, X-ray emission from clusters of galaxies. Cambridge Univ. Press, Cambridge
\bibitem{} Smail I., Couch W.J., Ellis R.S., Sharples R.M., 1995, ApJ, 440, 501
\bibitem{} Smail I., Ellis R.E., Dressler A., Couch W.J., Oemler A. Sharples R.M., Butcher H., 1997, ApJ, 479, 70
\bibitem{} Squires G., Kaiser N., Babul A., Fahlman G., Woods D., Neumann D.M., B\"ohringer H., 1996, ApJ, 461, 572
\bibitem{} Squires G., Neumann D.M., Kaiser N., Arnaud M., Babul A., B\"ohringer H., Fahlman G., Woods D., 1997, ApJ, 482, 648
\bibitem{} Thomas P.A., Muanwong O., Pearce F.R., Couchman H.M.P., Edge A.C., Jenkins A., Onuora L., 2000, MNRAS, submitted (astro-ph/0007348)
\bibitem{} van Haarlem M.P., Frenk C.S., White S.D.M., 1997, MNRAS, 287, 817 
\bibitem{} Waxman E., Miralda-Escud\'e J., 1995, ApJ, 451, 451
\bibitem{} Weisskopf M.C., Tananbaum H.D., Van Speybroeck L.P., O'Dell S.L., 2000, SPIE 4012, 1 (astro-ph/0004127)
\bibitem{} White D.A., 2000, MNRAS, 312, 663
\bibitem{} White D.A., Fabian A.C., 1995, MNRAS, 273, 72
\bibitem{} White D.A., Jones C., Forman W., 1997, MNRAS, 292, 419
\bibitem{} White S.D.M., Efstathiou G., Frenk C.S., 1993, MNRAS, 262, 1023
\bibitem{} Wise M.W., Sarazin C.L., 2000, ApJ, in press (astro-ph/9903119)
\bibitem{} Wu X., Fang L., 1997, ApJ, 483, 62 
\bibitem{} Wu X., 2000, MNRAS, 316, 299


\end{thebibliography}
\end{document}